\documentclass[iop]{emulateapj}

\usepackage{graphicx}
\usepackage{amssymb}
\usepackage{amsmath}
\usepackage{epstopdf}
\usepackage{color}

\def \deg {\ensuremath{^{\circ}}}
\def \arcsec {\ensuremath{^{\prime\prime}}}
\def \axp {4U\,0142+61}
\def \nustar {\emph{NuSTAR}}
\def \swift {\emph{Swift}}
\def \xmm {\emph{XMM-Newton}}

\newcommand{\AB}[1]{#1}
\newcommand{\revision}[1]{#1}

\newcommand{\alphamag}{\alpha_{\rm mag}}
\newcommand{\betaobs}{\beta_{\rm obs}}
\newcommand{\thetaj}{\theta_{\rm j}}
\newcommand{\phaseref}{\phi_{\rm 0}}

\newcommand{\lj}{L_j}
\newcommand{\fouru}{\axp}

\shorttitle{\nustar~ and \swift-XRT Observations of \axp}
\shortauthors{Tendulkar, S.~P., et al.}

\begin{document}

\setlength{\fboxsep}{0pt}%
\setlength{\fboxrule}{0.5pt}%

\title{Phase-resolved \emph{NuSTAR} and \emph{Swift}-XRT observations of Magnetar 4U\,0142+61}

\author{Shriharsh P. Tendulkar\altaffilmark{1}, Romain Hasc\"oet\altaffilmark{2}, Chengwei Yang\altaffilmark{3,4}, Victoria M. Kaspi\altaffilmark{4}, Andrei M. Beloborodov\altaffilmark{2}, Hongjun An\altaffilmark{5}, Matteo Bachetti\altaffilmark{6}, Steven E. Boggs\altaffilmark{7}, Finn E. Christensen\altaffilmark{8}, William W. Craig\altaffilmark{8,9}, Sebastien Guiilot\altaffilmark{4,10}, Charles A. Hailey\altaffilmark{2}, Fiona A. Harrison\altaffilmark{1}, Daniel Stern\altaffilmark{11}, William Zhang\altaffilmark{12}}
\email{spt@astro.caltech.edu}

\altaffiltext{1}{Space Radiation Laboratory, California Institute of Technology, 1200 E California Blvd, MC 249-17, Pasadena, CA 91125, USA}
\altaffiltext{2}{Columbia Astrophysics Laboratory, Columbia University, New York, NY 10027, USA}
\altaffiltext{3}{Beijing Institute of Technology, 5 South Zhongguancun Street, Haidian District, Beijing 100081,China}
\altaffiltext{4}{Department of Physics, McGill University, Montreal, Quebec, H3A 2T8, Canada}
\altaffiltext{5}{Kavli Institute for Particle Astrophysics and Cosmology, Stanford University, Stanford, CA 94305, USA}
\altaffiltext{6}{INAF/Osservatorio Astronomico di Cagliari, via della Scienza 5, I-09047 Selargius (CA), Italy
}
\altaffiltext{7}{Space Sciences Laboratory, University of California, Berkeley, CA 94720, USA}
\altaffiltext{8}{DTU Space, National Space Institute, Technical University of Denmark, Elektrovej 327, DK-2800 Lyngby, Denmark}
\altaffiltext{9}{Lawrence Livermore National Laboratory, Livermore, CA 94550, USA}
\altaffiltext{10}{Instituto de Astrof\'{i}sica, Facultad de F\'{i}sica, Pontificia Universidad Cat\'{o}lica de Chile, Av. Vicu\~{n}a Mackenna 4860, 782-0436 Macul, Santiago, Chile}
\altaffiltext{11}{Jet Propulsion Laboratory, California Institute of Technology, Pasadena, CA 91109, USA}
\altaffiltext{12}{NASA Goddard Space Flight Center, Astrophysics Science Division, Code 661, Greenbelt, MD 20771, USA}


\begin{abstract}
We present temporal and spectral analysis of simultaneous 0.5--79\,keV \swift-XRT and \nustar\ observations of the magnetar \axp. The pulse profile changes significantly with photon energy between 3 and 35\,keV. The pulse fraction increases with energy, reaching a value of $\approx$20\%, similar to that observed in 1E\,1841$-$045 and much lower than the $\approx$80\% pulse fraction observed in 1E\,2259+586.  We do not detect the 55-ks phase modulation reported in previous \textit{Suzaku}-HXD observations. The phase-averaged spectrum of \axp\ above 20\,keV is dominated by a hard power law with a photon index $\Gamma_H\sim0.65$, and the spectrum below 20\,keV can be described by two blackbodies, a blackbody plus a soft power law, or by a Comptonized blackbody model. We study the full phase-resolved spectra using the $e^\pm$ outflow model of Beloborodov (2013). Our results are consistent with the parameters of the active $j$-bundle derived from \textit{INTEGRAL} data by \citet{hascoet2014}. We find that a significant degeneracy appears in the inferred parameters if the footprint of the $j$-bundle is allowed to be a thin ring instead of a polar cap. The degeneracy is reduced when the footprint is required to be the hot spot inferred from the soft X-ray data.
\end{abstract}

\section{Introduction}
Magnetars --- isolated neutron stars with inferred surface \revision{dipolar} magnetic field strength, $B_\mathrm{surf} \gtrsim 10^{14}$\,G  --- were proposed to explain Soft Gamma Repeaters (SGRs) and later extended to include Anomalous X-ray Pulsars (AXPs)  \citep{thompson1995, thompson1996}. Unlike canonical pulsars powered by their rotational energy, the dominant energy reservoir of magnetars is their magnetic energy. The magnetar model attributes the anomalously high X-ray luminosity to the heating of the crust and magnetosphere by the dissipation of magnetic energy. There are 28 magnetars that have been discovered to date, including 5 candidates suggested on the basis of the detection of X-ray bursts \citep{olausen2014}\footnote{See the McGill Online Magnetar Catalog for a compilation of known magnetar properties: http://www.physics.mcgill.ca/$\sim$pulsar/magnetar/main.html}. \revision{Notably, two magnetars, SGR\,0418+5729 \citep{rea2010} and Swift\,J1822.3$-$1606 \citep[][and references therein]{scholz2014} were recently shown to have $B_\mathrm{surf} \sim 10^{12-13}$\,G, making them exceptions to the canonical classification above.} For recent reviews of magnetar observations we refer the readers to ~\citet{mereghetti2008} and \citet{rea2011}.

The X-ray to $\gamma$-ray spectrum of magnetars shows two peaks: a low-energy peak at $\sim$0.5\,keV and a high-energy peak at energies greater than $\sim$100\,keV \citep[see][and references therein]{kuiper2006, enoto2010}. The soft X-ray peak resembles a simple blackbody, likely originating from the magnetar surface, with a tail extending to $\sim$10\,keV caused by radiative transfer of photons through the magnetospheric plasma \citep[see for example][]{thompson2002}. Above energies of $\sim$10--20\,keV, the hard X-ray component is dominant.

The hard X-ray emission must be produced by the magnetosphere of the neutron star. A model making specific predictions for phase-resolved hard X-ray spectra emerged recently \citep{beloborodov2013a, beloborodov2013b} and was used to successfully fit the observations of  1E\,1841$-$045, 4U\,0142+61, 1RXS\,J1708$-$4009, and 1E\,2259+586 \citep{an2013,an2015,hascoet2014,vogel2014}.
 
The \emph{Nuclear Spectroscopic Telescope Array} (\emph{NuSTAR}) \citep{harrison2013} with its excellent spectral and timing capabilities is well-suited for phase-resolved spectroscopy of magnetars in the 3--79\,keV band (see \citealt{an2014} for a review). In addition to \axp\ {\em NuSTAR} has observed SGR\,1745$-$2900 \citep{mori2013,kaspi2014}, 1E\,1841$-$045 \citep{an2013,an2015}, 1E\,1048.1$-$5937 \citep{an2014b}, and 1E\,2259+586 \citep{vogel2014}.
 
\subsection{\axp}
\axp\ was discovered as a soft spectrum X-ray source in the \textit{Uhuru} all sky survey, initially reported in the analysis of the first 70 days of data \citep{giacconi1972}. It remained an unexceptional source until 8.7-s X-ray pulsations were discovered by \textit{ASCA} \citep{israel1993,israel1994}. 

The soft X-ray emission from \axp\ is well described by a blackbody with $k_\mathrm{B}T\sim 0.4$\,keV and a power law with index $\Gamma\sim 3.7$ \citep{white1996,israel1999,paul2000,juett2002,patel2003,gohler2004,gohler2005,rea2007,enoto2011}. Pulsed high-energy emission was detected between 20--50\,keV and 50--100\,keV using the IBIS/ISGRI instrument on \textit{INTEGRAL} \citep{denhartog2004ATel}. The hard X-ray spectrum ($>$10\,keV) is dominated by a power-law component with $\Gamma\sim1$ \citep{kuiper2006,denhartog2006,denhartog2008a,enoto2011}. Using upper limits on the $\gamma$-ray flux from the \textit{CGRO}-COMPTEL telescopes, the hard X-ray power-law cutoff energy was suggested to be between $\sim200$--$750$\,keV \citep{kuiper2006}. 

The soft X-ray pulse profiles of \axp\ were shown to undergo long-term changes using \emph{RXTE} observations spread over 10 years \citep{dib2007} and later \emph{Chandra}, \xmm\ and \swift\ observations \citep{gonzalez2010}. The pulse fractions were observed to increase over time leading up to a group of three bursts that occured between 2006--2007 \citep{gonzalez2010}. 

There has been much debate about the intrinsic soft X-ray spectra of magnetars, the measurement of which depends on the absorption column, $N_\mathrm{H}$, along the line of sight. \citet{durant2006b}, hereafter D06b, estimated the $N_\mathrm{H}$ to \axp\ to be $(6.4\pm0.7)\times10^{21}\,\mathrm{cm^{-2}}$ by fitting high-resolution grating spectra around individual photoelectric absorption edges of oxygen, iron, neon, magnesium and silicon. They also showed that the abundance ratios of Ne/Mg and O/Mg for \axp\ are closer to the revised solar abundances of \citet{asplund2005} compared to the old standard abundances of \citet{anders1989}. This measurement has the advantage of being less sensitive to the choice of model used to describe the intrinsic magnetar spectrum. However, since only the data near photo-electric edges is fitted, the fitting requires high-quality X-ray data. 

Unlike the $N_\mathrm{H}$ measurements from the high-resolution X-ray spectra, most fits of the low-energy spectrum ($\approx$ 0.5--10\,keV) with a blackbody plus power-law model converge to $N_\mathrm{H}\approx 1.0\times10^{22}\,\mathrm{cm^{-2}}$ \citep[][and references therein]{rea2007} which used the abundances from \citet{anders1989}. We note that \citet{enoto2011} obtained  $N_\mathrm{H}$ values consistent with the D06b value from broadband spectral fits to \emph{Suzaku} data, however, no abundance model was specified. In this work, we use solar abundance values from \citet{asplund2009} --- the `\texttt{aspl}' model in \texttt{XSPEC} --- as default, and we also test our fits with other abundance models.

 By identifying core helium-burning giant stars --- i.e. red clump stars --- from the 2MASS catalog and estimating the variation of optical extinction as a function of distance in the direction of magnetars, \citet{durant2006a} estimated the distance to \axp~ to be $3.6\pm0.4$\,kpc. This distance estimate used the $N_\mathrm{H}=(6.4\pm0.7)\times10^{21}\,\mathrm{cm^{-2}}$ estimated from the photo-electric absorption edges. Previous measurements of optical extinction ($A_V$) have concluded that the $A_V$ for \axp\ should be less than 5 \citep{hulleman2004, durant2006b}, corresponding to $N_\mathrm{H}<9\times10^{21}\,\mathrm{cm^{-2}}$ \citep[$A_V=N_\mathrm{H}\times5.6\times10^{22}\,\mathrm{cm^{2}}$]{predehl1995}.

In this paper, we present a phase-resolved spectral and timing analysis of coordinated \swift-XRT and \nustar\ spectra of \axp. We use the \citet{hascoet2014} framework to test the electron-positron outflow model and constrain physical parameters. The paper is organized as follows. In Section~\ref{sec:obs}, we describe our observations, data and data reduction procedure. In Section~\ref{sec:results}, we describe the results of our timing analysis, spectral analysis and model fitting. In Section~\ref{sec:discussion}, we discuss these results in the context of previous observations of \axp\ and those of other magnetars.

\section{Observations and Analysis}
\label{sec:obs}
\nustar\ \citep{harrison2013} is a 3--79\,keV focusing hard X-ray mission. It consists of two identical co-aligned Wolter-I telescopes with CdZnTe detectors at the focal planes. The telescopes provide a point-spread function with a half-power diameter (HPD) of 58\arcsec\ over a field of view of 12\arcmin$\times$12\arcmin. The energy resolution varies from 0.4\,keV at 6\,keV to 0.9\,keV at 60\,keV. The two focal plane modules are referred to as FPMA and FPMB. \axp\ was observed by \nustar\ between 2014 Mar 27 -- 30 during a 44-ks observation simultaneous with a 24-ks observation with the \swift\ X-ray Telescope~\citep[XRT; ][]{burrows2005}. The details of the observations are summarized in Table~\ref{tab:obs}.

We performed the processing and filtering of the \nustar\ event data with the standard \nustar\ pipeline version 1.4.1 and \texttt{HEASOFT} version 6.16. We used the \texttt{barycorr} tool to correct the photon arrival times for the orbital motion of the satellite and the Earth at the optical position of \axp\ --- $\alpha=01^\mathrm{h}46^\mathrm{m}22^\mathrm{s}.407,~\delta=+61\deg45\arcmin03\arcsec.19$ (J2000) --- as reported by \citet{hulleman2004}. The source events were extracted within a 50-pixel (120\arcsec) radius around the centroid and suitable background regions were used. Spectra were extracted using the \texttt{nuproducts} script. Using \texttt{grppha}, all photons below channel 35 (3\,keV) and above channel 1935 (79\,keV) were flagged as bad and all good photons were binned in energy to achieve a minimum of 30 photons per bin. 

The \swift-XRT data \revision{were obtained in the Windowed Timing (\texttt{WT}) mode and} were processed with the standard \texttt{xrtpipeline} and the photon arrival times were corrected using \texttt{barycorr}. The \texttt{xrtproducts} script was used to extract spectra and lightcurves within a radius of 25\,pixels (59\arcsec)\revision{\footnote{As per the \swift-XRT data analysis thread: \url{http://www.swift.ac.uk/analysis/xrt/spectra.php}.}}. Photons in channels 0--29 (energy $<0.3$\,keV) were ignored and all channels between 0.3--10\,keV were binned to ensure a minimum of 30 photons per bin. 

The \swift-XRT and \nustar\ FPMA and FPMB spectra were fit simultaneously in \texttt{XSPEC} v12.8.1 \citep{arnaud1996} using two freely varying cross-normalization constants, assuming the normalization of \swift-XRT to be fixed to unity. Timing analysis was performed on exposure-corrected lightcurves and event lists using custom \texttt{MATLAB} scripts. 

\begin{deluxetable}{lrrrr}
  \centering
  \tablecolumns{5} 
  \tablecaption{X-ray observations of \axp\ in 2014.\label{tab:obs}}
  \tablewidth{0pt}
  \tabletypesize{\footnotesize}
  \tablehead{
    \colhead{Obs ID}   &
    \colhead{Start}      &
    \colhead{End}   &
    \colhead{Exp}      &
    \colhead{Rate\tablenotemark{a}}\\
    \colhead{}  &
    \colhead{(UT)}                    &
    \colhead{(UT)}                    &
    \colhead{(ks)}                 &
    \colhead{cts/s}
  }
  \startdata
  \sidehead{\nustar}
  30001023002 & Mar 27 13:35 & Mar 28 00:45 & 7  & 1.3\\
  30001023003 & Mar 28 00:45 & Mar 30 13:00 & 37 & 1.3\\
  \sidehead{\swift-XRT (Windowed Timing Mode)}
  00080026001 & Mar 27 13:36 & Mar 27 21:52 & 4.9  & 4.2\\
  00080026002 & Mar 28 07:10 & Mar 29 23:15 & 12.9 & 4.0\\
  00080026003 & Mar 30 00:42 & Mar 30 08:57 & 6.6  & 4.1
  \enddata
\tablenotetext{a}{0.5--10\,keV count rate for \swift-XRT and 3--79\,keV count rate from \nustar.}
\end{deluxetable}

\section{Results}
\label{sec:results}

\subsection{Pulse Profile}
\label{sec:pulse_profile}
We analyzed the barycentered 3--79\,keV \nustar\ events using epoch folding \citep{leahy1987} and measured the rotation period of \axp\ to be $P=8.689158(4)$\,s. \revision{This is consistent with the period measured with the \swift-XRT observation and is also} consistent with the period $P=8.689163(5)$\,s expected at the epoch of observation based on the last ephemeris measured after the glitch of 2011 July, reported by \citet{dib2014}.

\begin{figure*}
  \center
  \includegraphics[clip=true,trim=0.2in 0.00in 0.6in 0.4in,width=0.32\textwidth]{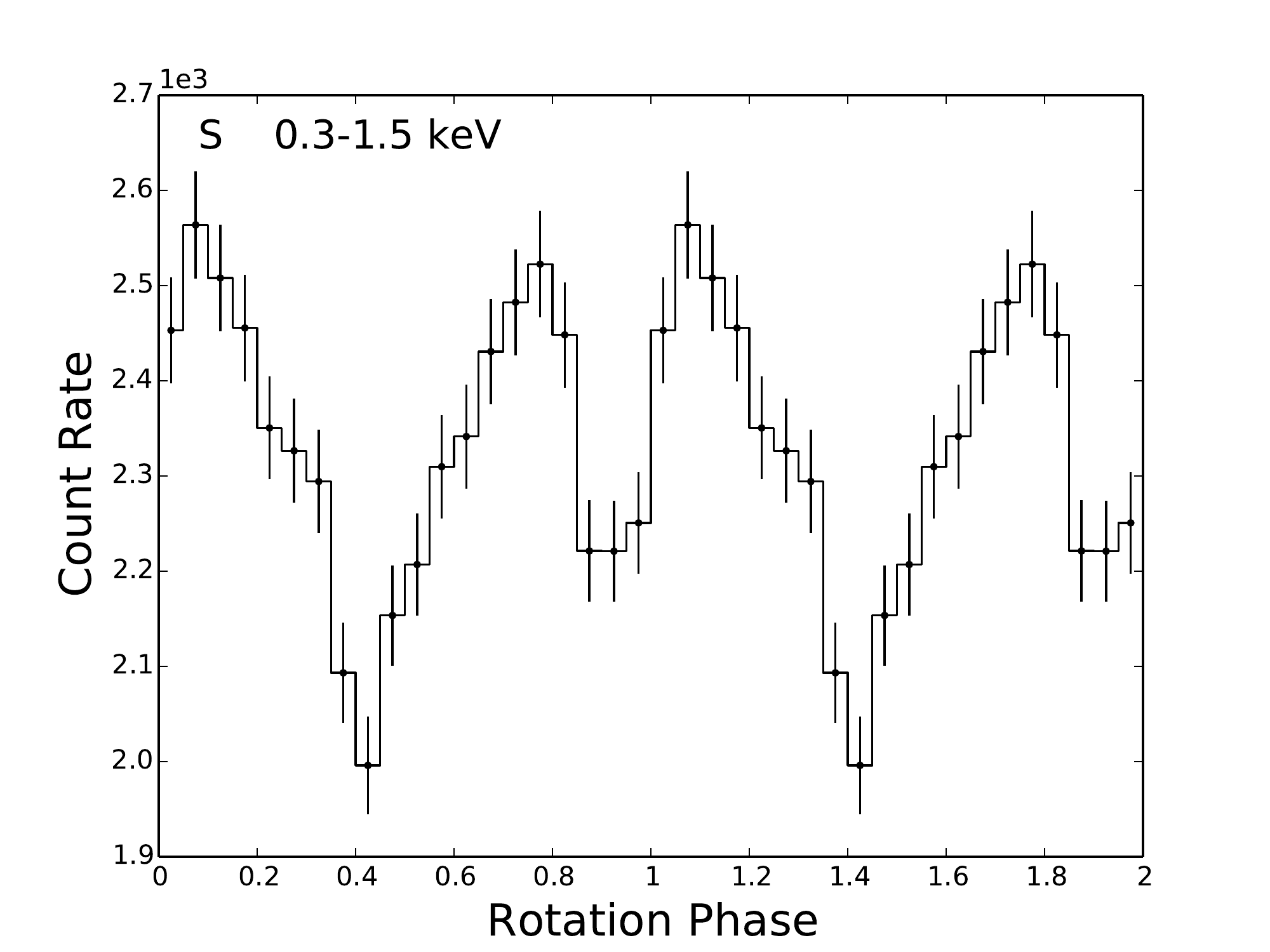}
  \includegraphics[clip=true,trim=0.2in 0.00in 0.6in 0.4in,width=0.32\textwidth]{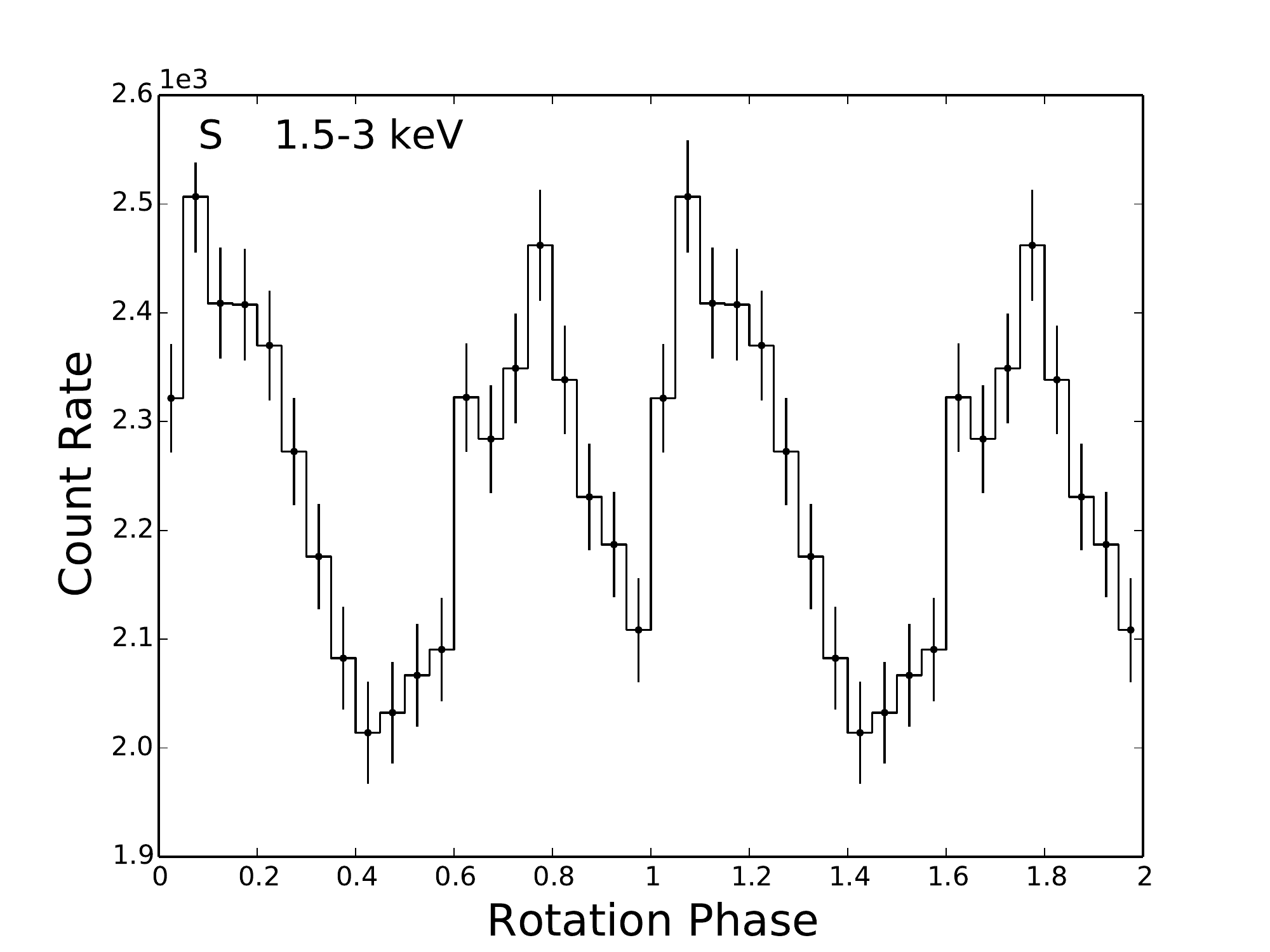}
  \includegraphics[clip=true,trim=0.2in 0.00in 0.6in 0.4in,width=0.32\textwidth]{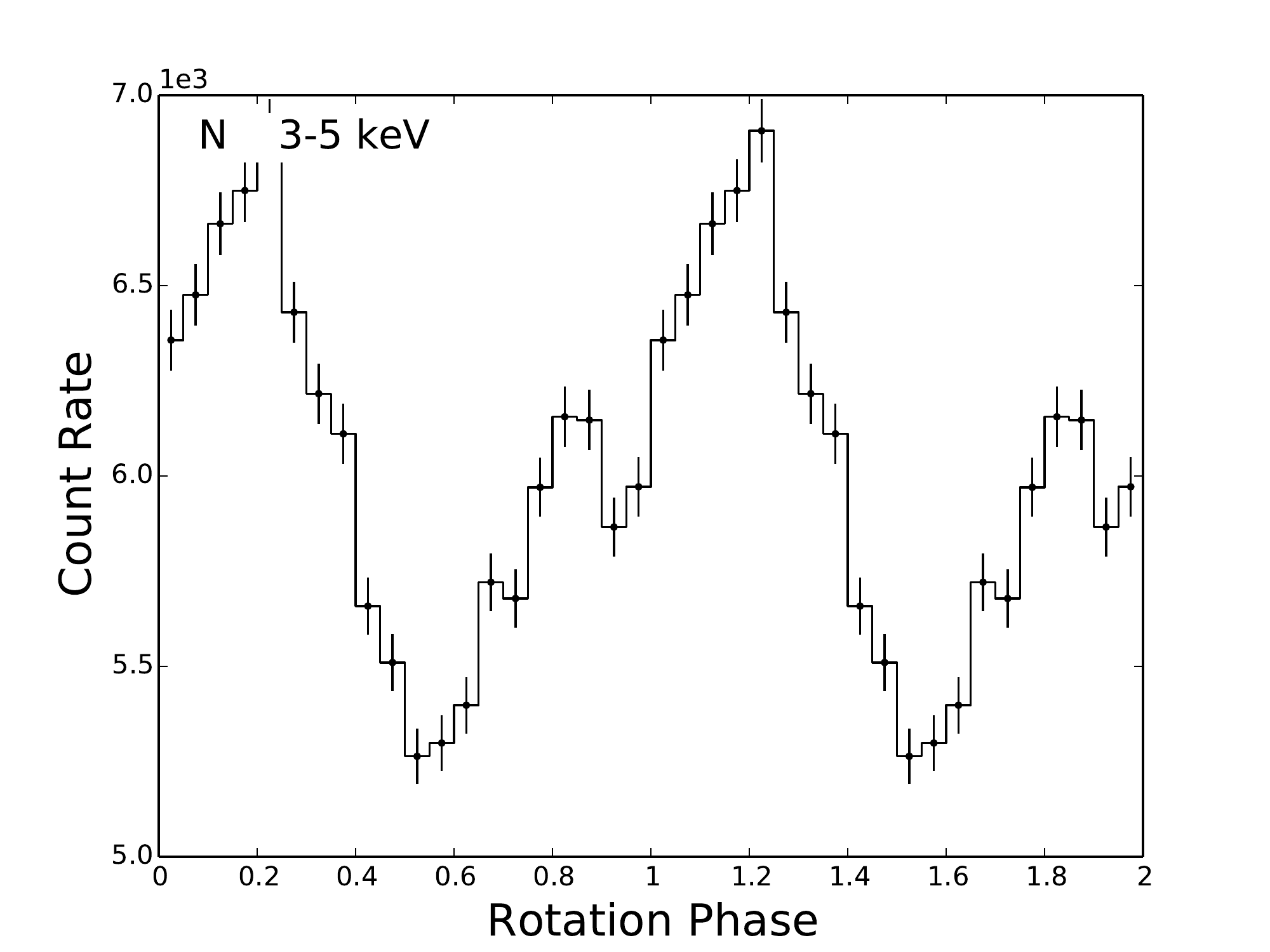}
  \includegraphics[clip=true,trim=0.2in 0.00in 0.6in 0.4in,width=0.32\textwidth]{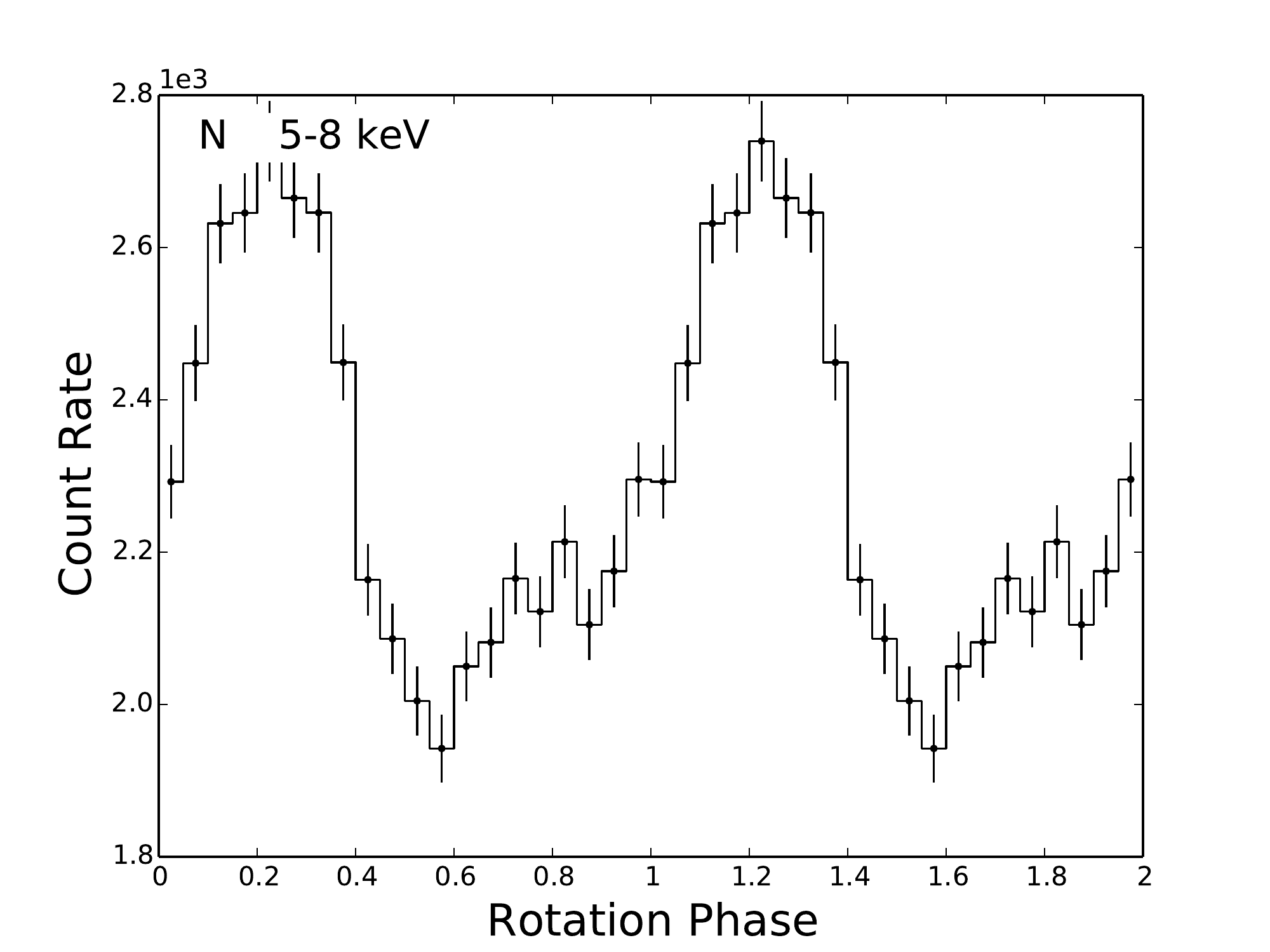}
  \includegraphics[clip=true,trim=0.2in 0.00in 0.6in 0.4in,width=0.32\textwidth]{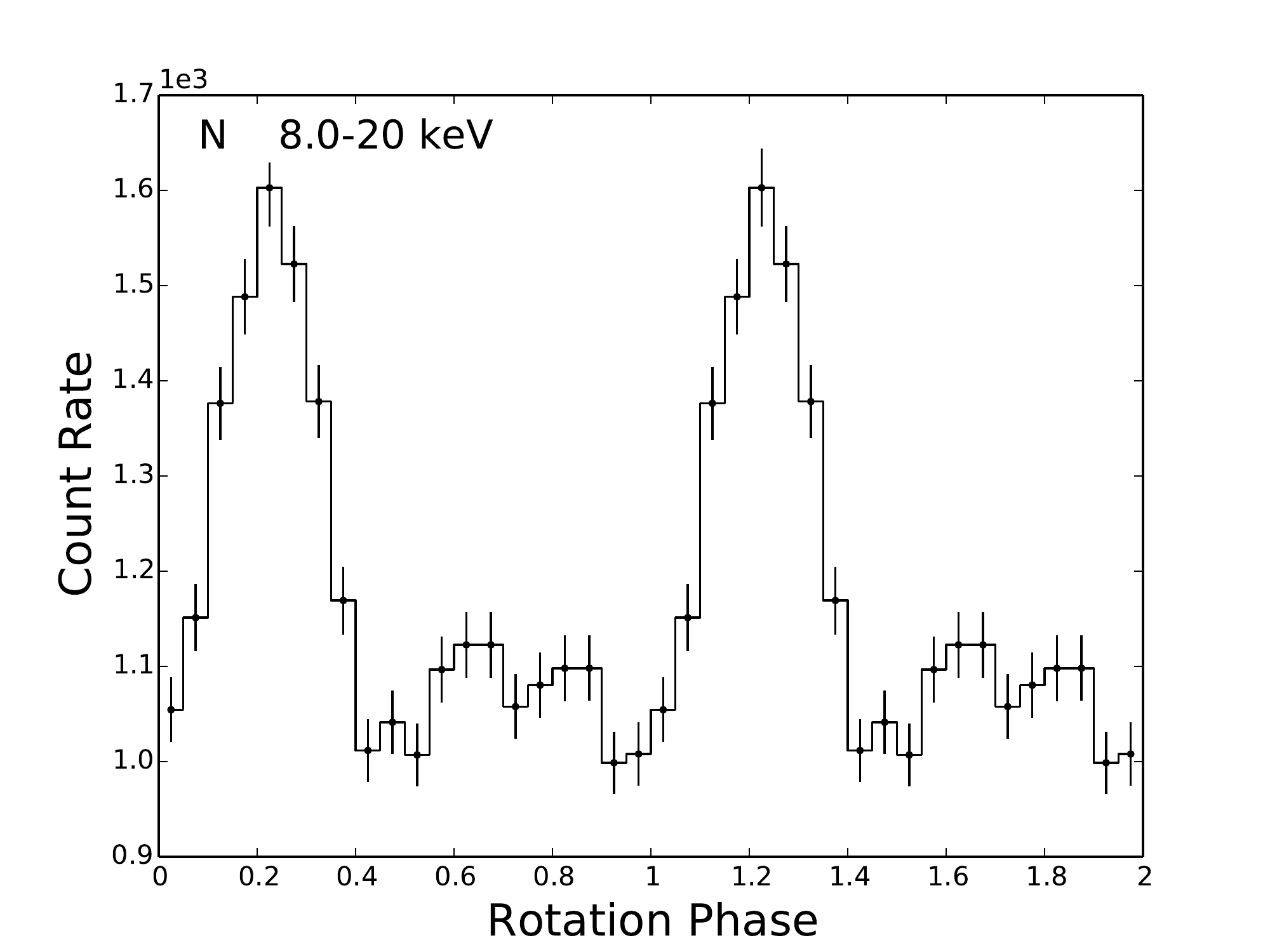}
  \includegraphics[clip=true,trim=0.2in 0.00in 0.6in 0.4in,width=0.32\textwidth]{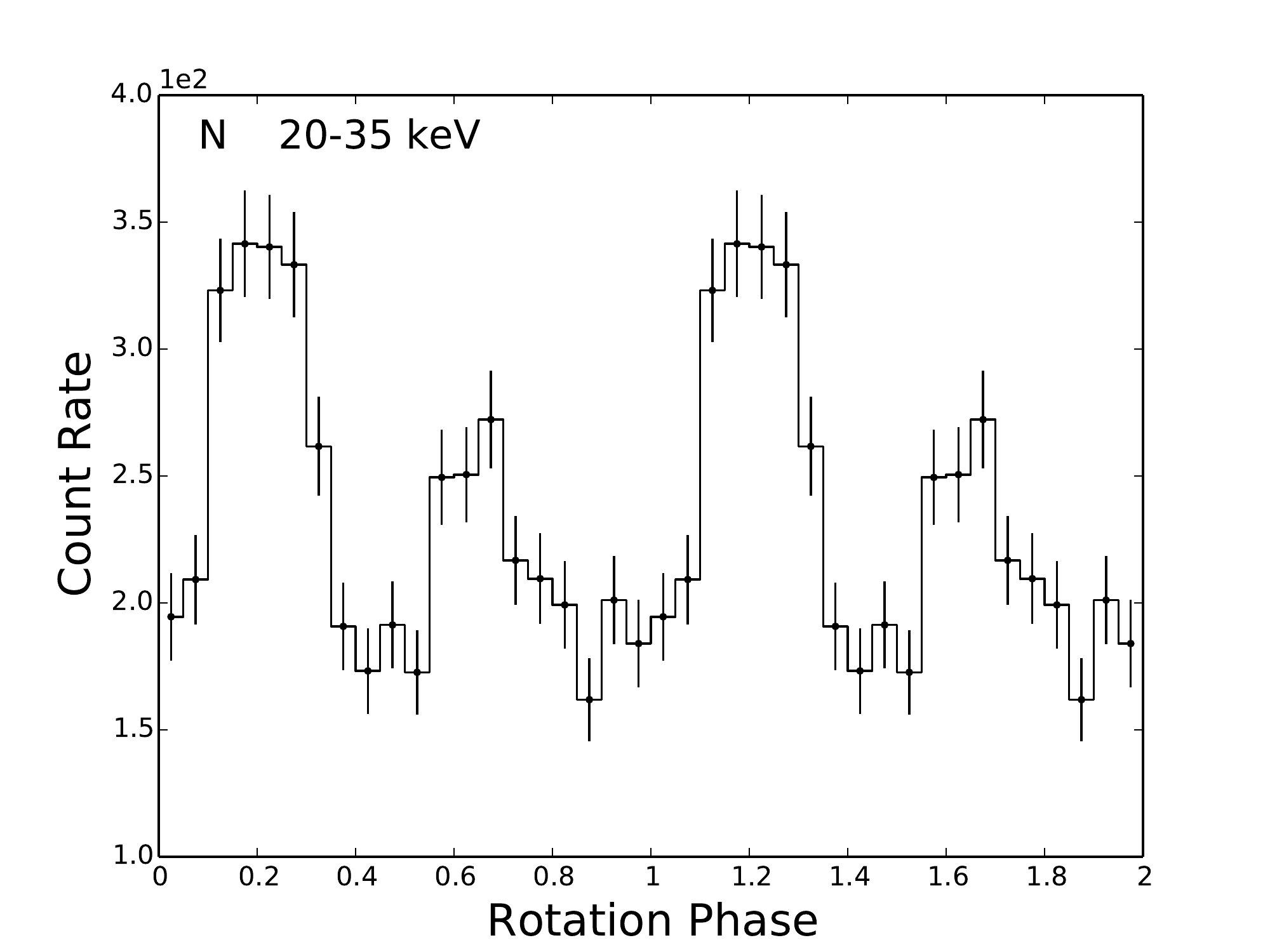}
  \includegraphics[clip=true,trim=0.2in 0.00in 0.6in 0.4in,width=0.32\textwidth]{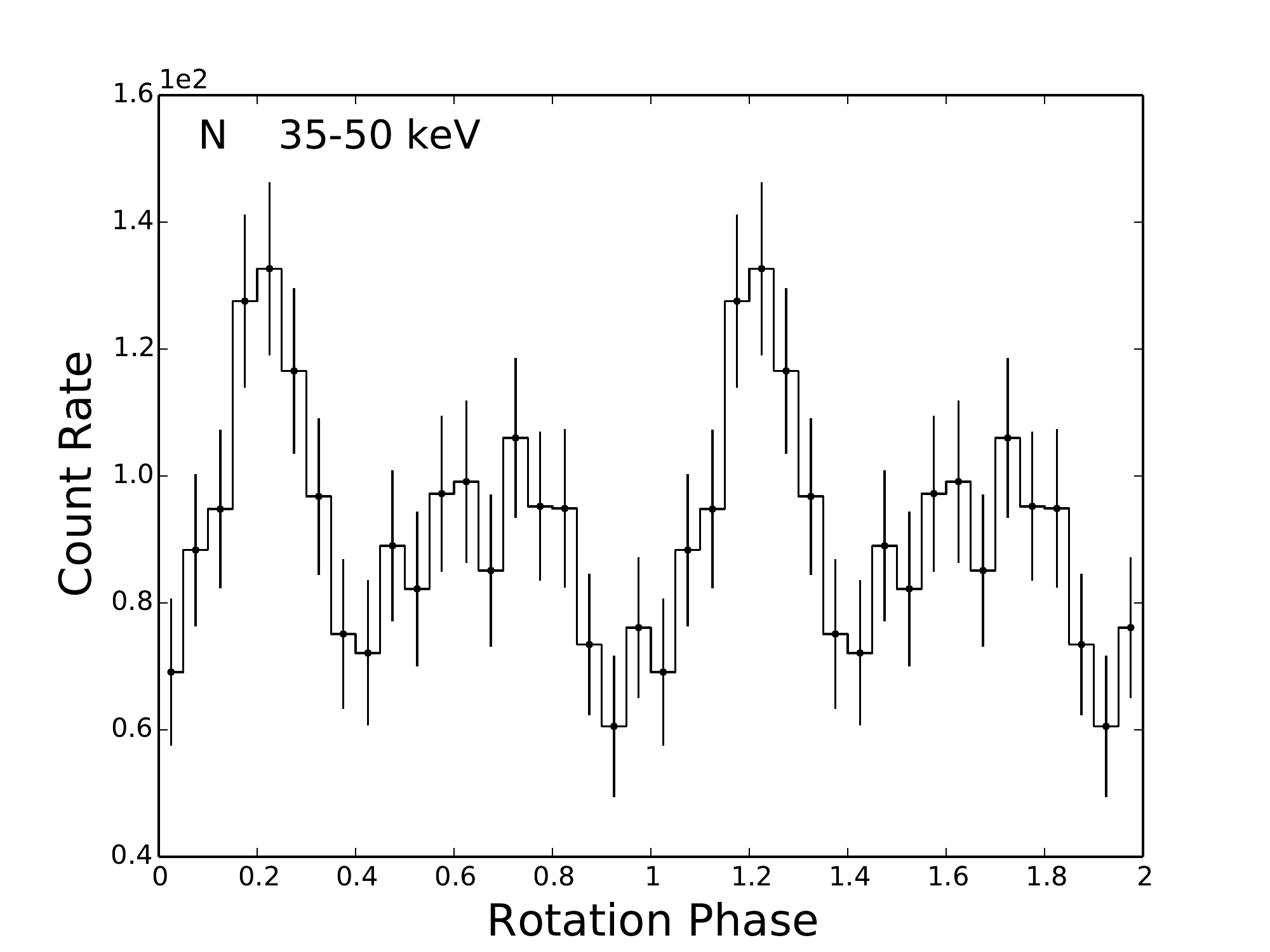}
  \includegraphics[clip=true,trim=0.2in 0.00in 0.6in 0.4in,width=0.32\textwidth]{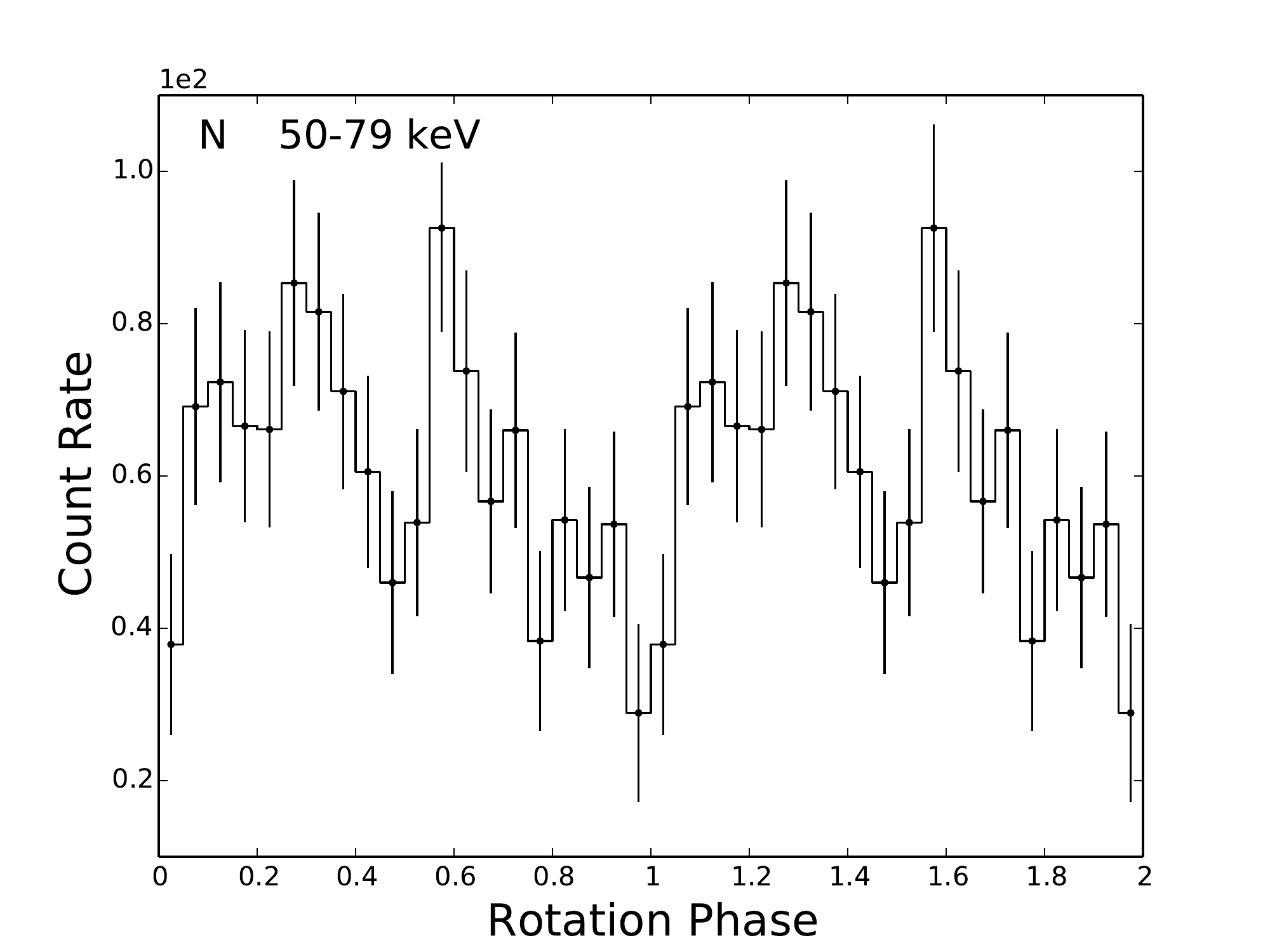}
  \includegraphics[clip=true,trim=0.2in 0.00in 0.6in 0.4in,width=0.32\textwidth]{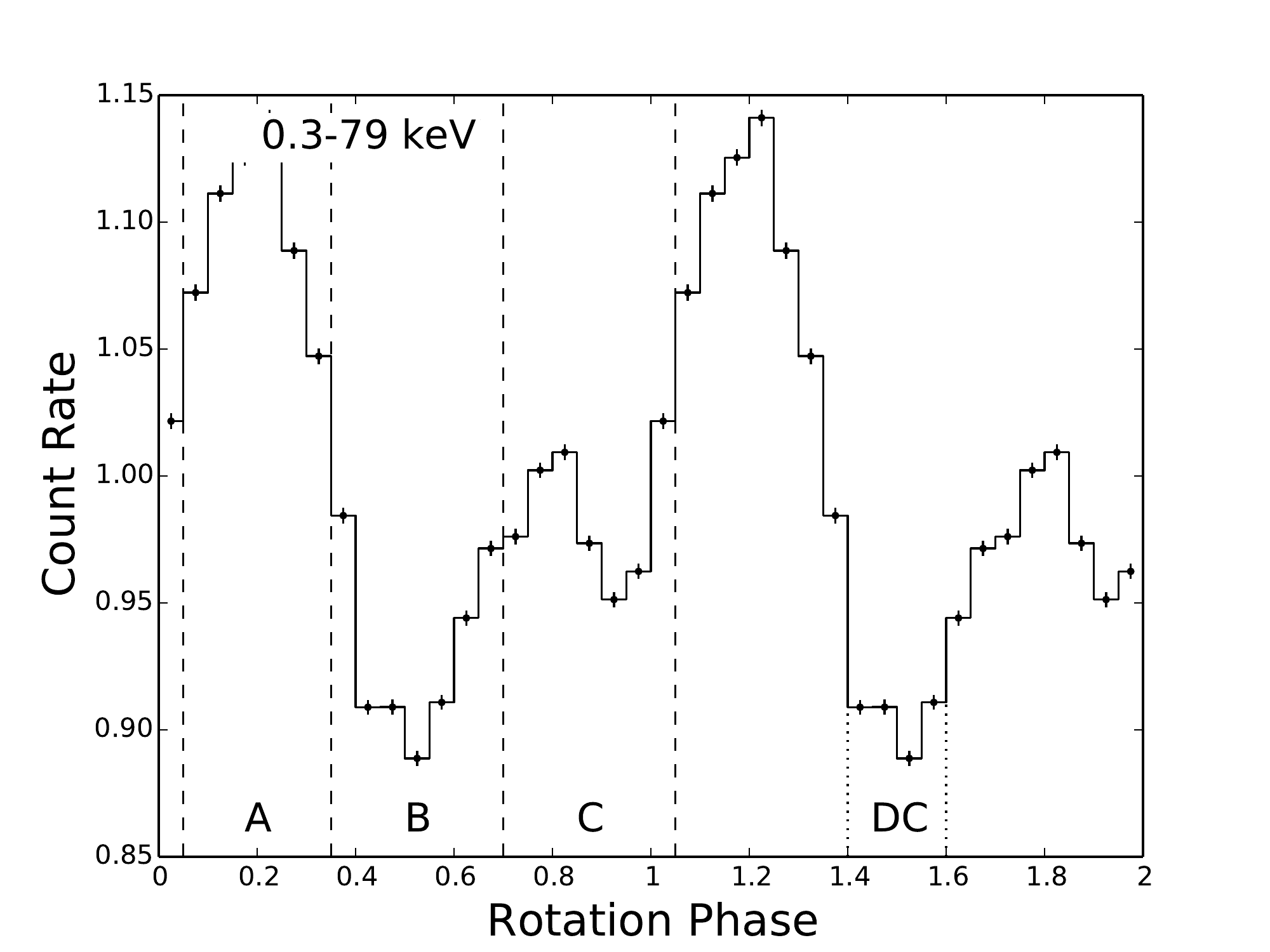} 
  \caption{\swift-XRT and \nustar\ pulse profiles in different energy bands. The annotation in the upper left corner of each plot specifies the telescope (`S': \swift-XRT, `N': \nustar) and the energy band for each plot. The last plot is the total 0.3--79\,keV (\swift-XRT and \nustar) count rate (normalized to the average value) marked with phase bins --- `A', `B' and `C' and `DC' --- used for fitting the $e^{-}-e^{+}$ outflow model (Section~\ref{sec:model_fits}). Two pulse periods are shown for clarity.}
  \label{fig:pulse_profile}
\end{figure*}

We folded the \swift-XRT and \nustar\ events in 8 energy bands into 20 phase bins with our measured period to compare the pulse morphology as a function of energy (Figure~\ref{fig:pulse_profile}). The energy bands --- 0.3--1.5\,keV, 1.5--3\,keV, 3--5\,keV, 5--8\,keV, 8--20\,keV, 20--35\,keV, 35--50\,keV and 50--79\,keV --- were chosen to have approximately equal counts in each band. The 3--5\,keV and 5--8\,keV data from \swift-XRT had far lower count rates than the corresponding \nustar\ observations and hence were not used for the final analysis. However, we confirmed that the \nustar\ and \swift-XRT pulse profiles in these two overlapping energy bands are consistent within the error bars. 

There is a clear gradual change in the pulse morphology as the energy band crosses $\sim$3\,keV and $\sim$20\,keV, corresponding to the different spectral components --- modified blackbody or hard power law --- that dominate the spectrum at these energies. This change in morphology is also observed in the dominance of the Fourier harmonics described in Section~\ref{sec:morphology}. The 0.3--1.5\,keV and 1.5--3\,keV pulse profiles consist of two peaks at phases of $\phi=0.3$ and $\phi=0.6$ separated by a sharp dip at $\phi=0.5$. In the 3--5\,keV and 5--8\,keV bands, the peak at $\phi=0.3$ dominates the pulse and the dip at $\phi=0.5$ deepens significantly. There is a small dip at $\phi=0.9$ separating a possible second pulse peak from the primary. Moving to higher photon energies, in the  8--20\,keV and 20--35\,keV bands, a second pulse rises in amplitude at $\phi\approx0.65$ towards energies of 50\,keV. The primary pulse also shows signs of broadening as a function of energy.

Our low-energy results are consistent with the \emph{RXTE} observations reported by \citet{denhartog2008a} (hereafter dH08) and with the 0.5--10\,keV \xmm\ observations of \citet{gonzalez2010} from 2008 March. However, the 20--35\,keV and 35--50\,keV observations from \nustar\ show a double peak structure with the primary peak having approximately twice the peak amplitude as compared to the secondary peak. The corresponding \emph{INTEGRAL} pulse profile reported in dH08 showed a double peak structure with both peaks of equal amplitude. This difference is also present in the pulse fraction analysis presented in Section~\ref{sec:pulse_fraction}. 

\subsection{Pulse Morphology}
\label{sec:morphology}
To explore the variation in pulse shape as a function of energy, we decomposed the pulses into Fourier harmonics. We define Fourier coefficients $a_k$ and $b_k$ as
\begin{equation}
  a_k = \frac{1}{N}\sum_{j=1}^{N}p_j \cos\left(\frac{2\pi k j}{N}\right)~\mathrm{and}
\end{equation}
\begin{equation}
  b_k = \frac{1}{N}\sum_{j=1}^{N}p_j \sin\left(\frac{2\pi k j}{N}\right)
\end{equation}
where $N$ is the number of phase bins and $p_j$ is the number of photons in each phase bin \revision{ and $j$ and $k$ are indices referring to the phase bins and the Fourier harmonics respectively}. We define the strength of each Fourier component to be $A_k=\sqrt{a_k^2+b_k^2}$. We define $A_\mathrm{total}$ as
\begin{equation}
  A_\mathrm{total}=\sqrt{\sum_{k=1}^N A_k^2}.
\end{equation} 

\begin{figure}
  \center
  \includegraphics[clip=true,trim=0.2in 0.05in 0.6in 0.5in,width=0.48\textwidth]{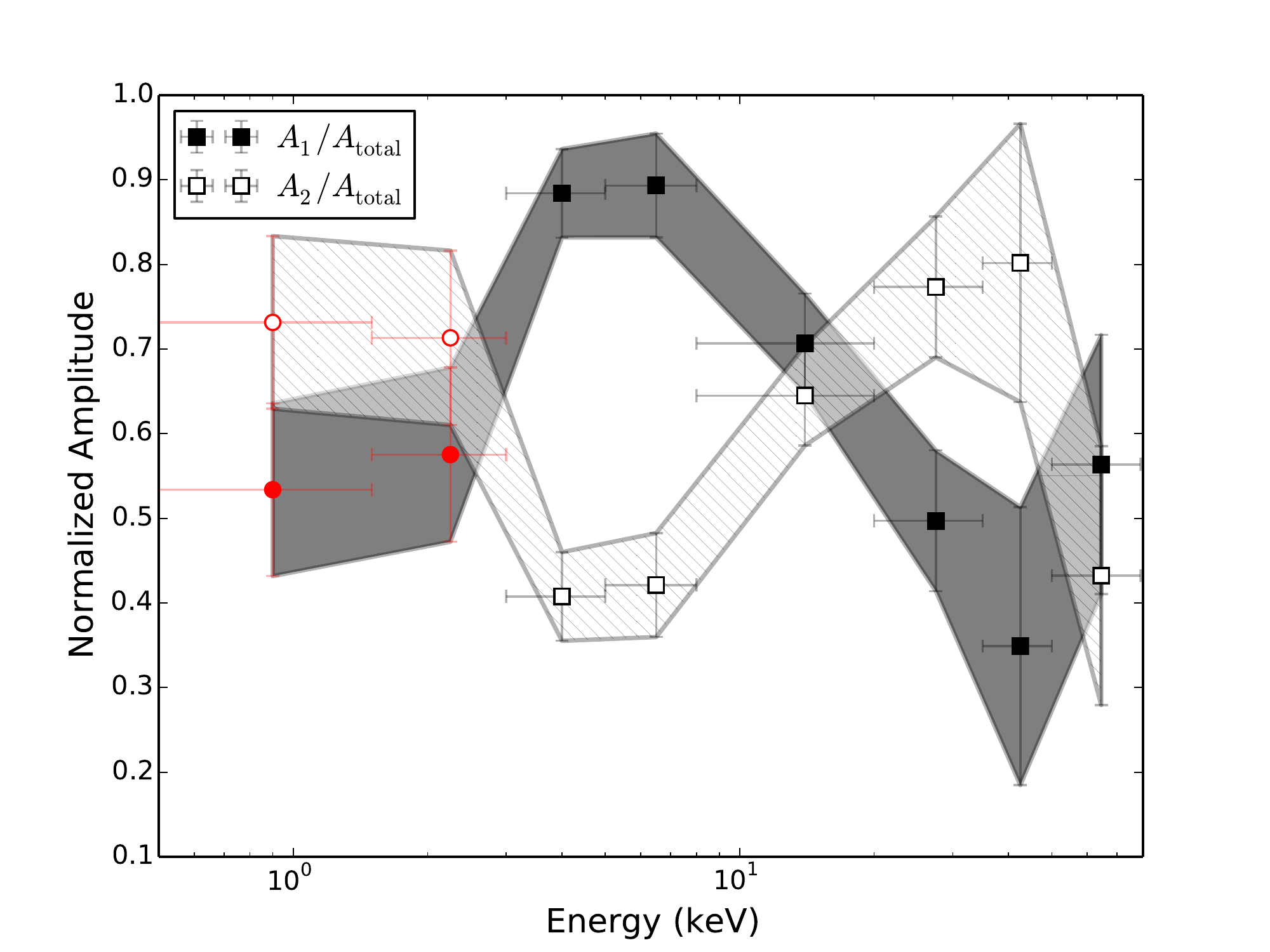}
  \caption{Variation in the first (filled symbols) and second (empty symbols) harmonic amplitudes as a function of photon energy. Both values are normalized with respect to the total amplitude of the variation ($A_\mathrm{total}$). The \swift-XRT data points are shown as red circles and the \nustar\ data points are shown as black squares. The filled areas (solid for $A_1/A_\mathrm{total}$ and hashed for $A_2/A_\mathrm{total}$) show the 1-$\sigma$ error regions. (A color version of this figure is available in the online journal.)}
  \label{fig:harmonic_variation}
\end{figure}

We find that most of the variational power in the pulses is explained in the first six harmonic coefficients. The distinct variation of pulse shapes with energy can be seen in Figure~\ref{fig:harmonic_variation}. The fraction of power in the first harmonic ($A_1/A_\mathrm{total}$) decreases with energy until approximately 40\,keV and then increases, whereas the fraction of power in the second harmonic ($A_2/A_\mathrm{total}$) increases with energy until approximately $40$\,keV and then decreases.

This behavior of the harmonics is significantly different from that of 1E\,2259+586 presented in \citet{vogel2014}. In 1E\,2259+586, the normalized $A_1$ value increases as a function of energy until approximately 12\,keV and then decreases as a function of energy. The normalized value of $A_2$ decreases as a function of energy until approximately 12--15\,keV and then increases.

\subsection{Pulse Fraction}
\label{sec:pulse_fraction}
We quantify the strength of the pulsations using two different methods. We define the root-mean-square pulse fraction as
\begin{equation}
PF_\mathrm{RMS}=\frac{\sqrt{2\sum_{k=1}^{N}((a_k^2+b_k^2)-(\sigma_{a_k}^2+\sigma_{b_k}^2))}}{a_0},
\end{equation}
where $a_k$ and $b_k$ are as defined above and $\sigma_{a_k}$ and $\sigma_{b_k}$ are the uncertainties in $a_k$ and $b_k$, respectively, calculated using Poisson variances as
\begin{equation}
\sigma_{a_k}^2 = \frac{1}{N^2}\sum_{j=1}^{N}\sigma^2_{p_j} \cos^2\left(\frac{2\pi k j}{N}\right),
\end{equation}
\begin{equation}
\sigma_{b_k}^2 = \frac{1}{N^2}\sum_{j=1}^{N}\sigma^2_{p_j} \sin^2\left(\frac{2\pi k j}{N}\right).
\end{equation}
This definition, including the correction term, $\sigma_{a_k}^2+\sigma_{b_k}^2$, has been shown to be a robust and accurate metric of pulse fraction in noisy data \citep[see Appendix 1 of ][ for a detailed discussion]{an2015}. 

We also define the area pulse fraction described by \citet{gonzalez2010} as
\begin{equation}
PF_\mathrm{area}=\frac{\sum_{j=1}^{N}p_j-N*\min(p_j)}{\sum_{j=1}^{N}p_j}.
\end{equation}
This definition is consistent with that used by dH08. However, it is challenging to determine the true value of $\min(p_j)$, and both noise and binning tend to bias the $PF_\mathrm{area}$ metric upwards by as much as 20\% \citep{an2015}. 

\begin{figure}
\center
\includegraphics[clip=true,trim=0.2in 0.05in 0.6in 0.5in,width=0.48\textwidth]{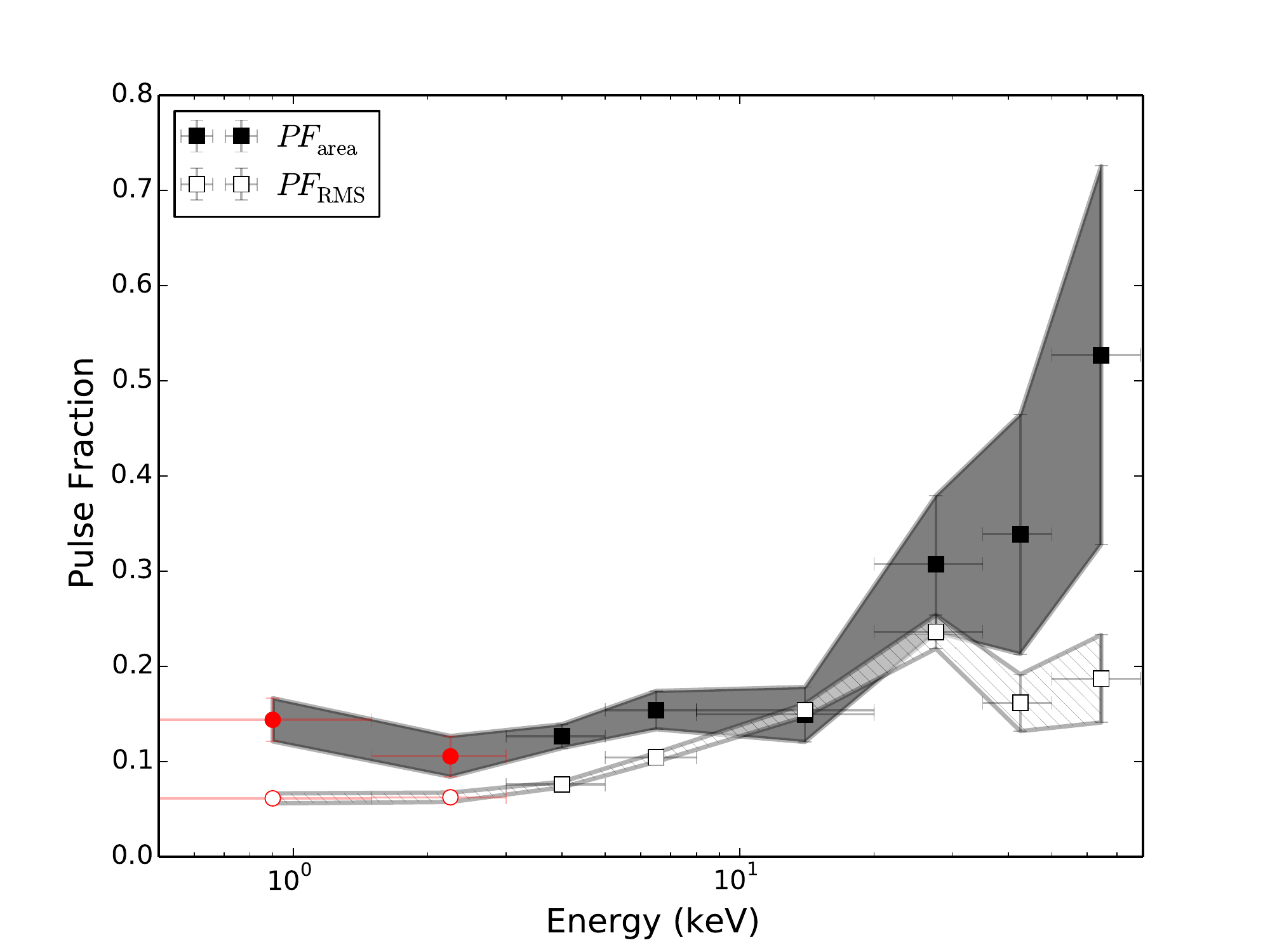}
\caption{Variation in the area pulsed fraction (filled symbols) and root-mean-square pulsed fraction (empty symbols) as a function of energy. The \swift-XRT and \nustar\ symbols are the same as in Figure~\ref{fig:harmonic_variation}. The filled areas (solid for $PF\mathrm{area}$ and hashed for $PF_\mathrm{RMS}$) show the 1-$\sigma$ error regions. A color version of this figure is available in the online journal.}
\label{fig:pulse_fraction}
\end{figure}

Figure~\ref{fig:pulse_fraction} shows the variation of $PF_\mathrm{area}$ (filled symbols) and $PF_\mathrm{RMS}$ (empty symbols) as a function of energy. Note that while our measurements of $PF_\mathrm{area}$ have an increasing trend at energies $>$10\,keV, the $PF_\mathrm{area}$ values above 20\,keV are also consistent with a constant value of $\approx$35\%.  The near-linear increase in $PF_\mathrm{area}$ as a function of energy is consistent with the results of dH08, though we note that our $PF_\mathrm{area}$ measurements are consistently higher than those of dH08 and those of \citet{gonzalez2010}. The RMS pulse fraction, $PF_\mathrm{RMS}$, increases with energy up to an energy of 35\,keV. However, the absolute normalization is different due to the different definitions of pulse fractions. The possible decrease in $PF_\mathrm{RMS}$ in the 35--50\,keV and 50--79\,keV bands may be due to the emergence of two nearly equal amplitude peaks in the pulse profile with lower count rates. A similar reduction in RMS pulsed fraction with energy had been reported for 1E\,1841$-$045 in the 16--24\,keV band observations \citep{an2013}. However, with more \nustar\ observations the variations were shown to be dependent on the exact energy bins used \citep{an2015}. The overall trend of both RMS and area pulse fraction was shown to increase with energy, with $PF_\mathrm{RMS}$ increasing up to 20\% at 50\,keV and $PF_\mathrm{area}$ increasing to 50\% at 50\,keV. Similar to 1E\,1841$-$045, $PF_\mathrm{RMS}$ does not show signs of increasing to 100\% with increasing energy as was suggested from the \emph{INTEGRAL} data (dH08). 

\subsection{Non-Detection of Precession}
\citet{makishima2014} (hereafter ME14) reported a phase modulation in the 8.7-s rotation period of \axp\ with an amplitude of 0.7\,s and a period of 55$\pm4$\,ks ($\approx$15\,hr) detected from 15--40\,keV HXD-PIN data gathered with \textit{Suzaku} in 2009 August. This was interpreted as possible evidence for the precession of the neutron star caused by slight deviation from spherical symmetry. The same search in \emph{Suzaku}-HXD data gathered in 2007 August and XIS data from 2007 August and 2009 August did not lead to detection of precession. Since our observations are spread over 4\,days, we searched for possible variations in the rotation period or rotation phase following the same $Z^2_n$ analysis \citep{brazier1994} steps reported by ME14.

\begin{figure}
\center
\includegraphics[clip=true,trim=0.2in 0.1in 0.6in 0.35in,width=0.48\textwidth]{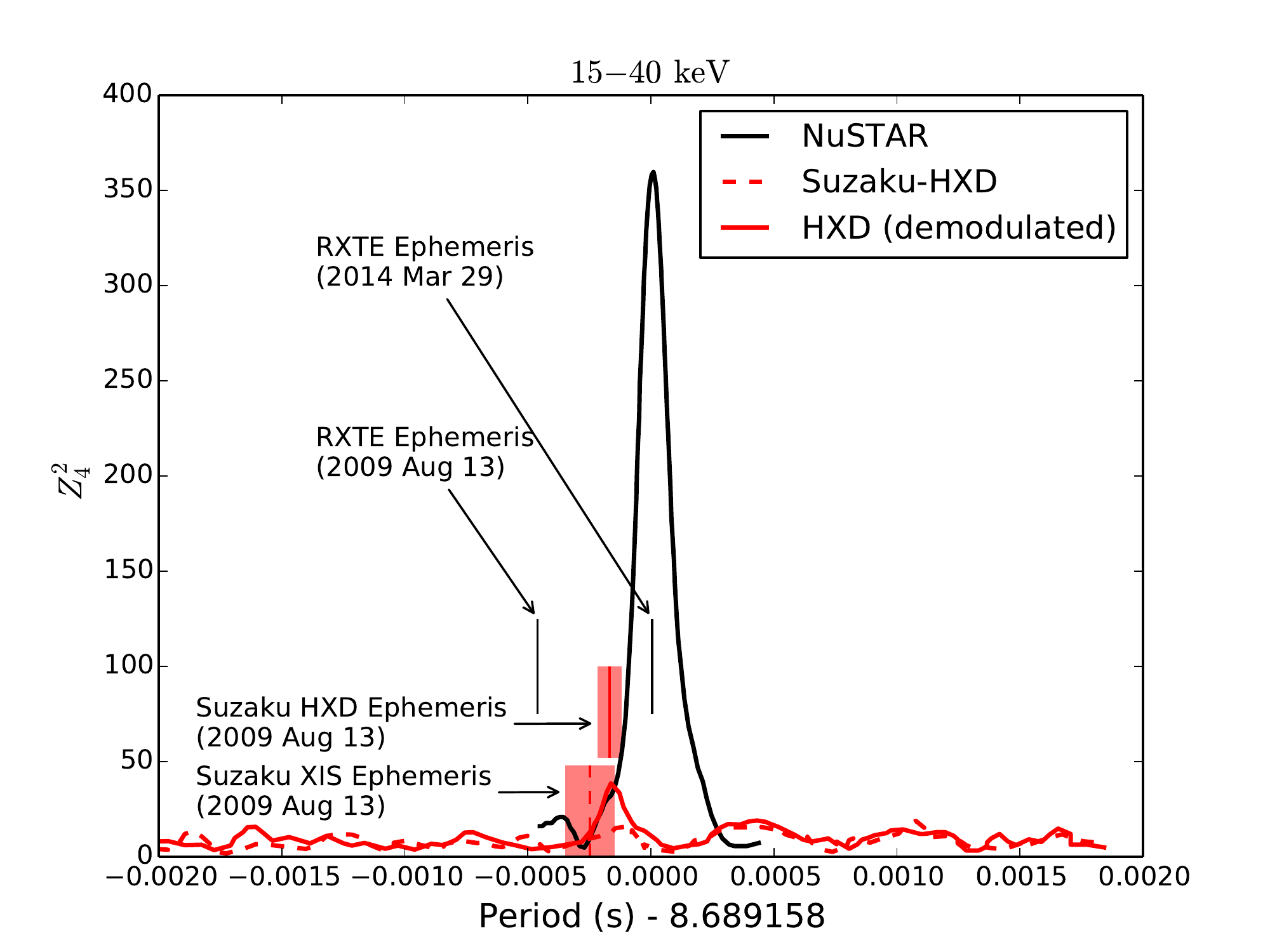}
\caption{Variation of $Z^2_4$ as a function of rotation period for \axp\ for 15--40\,keV band using \nustar\ data (solid black curve) overlaid on \emph{Suzaku}-HXD results from ME14 (red dashed and solid curves). We find that $Z^2_4$ peaks to a value of $\approx$360 at the measured rotation period of 8.689158\,s (Section~\ref{sec:pulse_profile}). The dotted red line is $Z^2_4$ for the raw (non-demodulated) \textit{Suzaku}-HXD data reported in Figure 1b of ME14. The solid red line is $Z^2_4$ for the same HXD data after optimal demodulation. The vertical dashed red line is the high-energy (15--40\,keV) rotational period reported by ME14 with the shaded region denoting the reported error and the vertical dashed red line the low-energy (0.3--10\,keV) rotational period reported from the XIS data for the same epoch. Vertical black lines are the expected rotation periods for the 2009 August and 2014 March epochs from the \emph{RXTE} data. The rotation periods marks are offset vertically for clarity.}
\label{fig:z4_makishima}
\end{figure}

For $n=3$ and $n=4$, we find that $Z^2_n$ peaks to a value of $\approx$360 at the rotation period of $P=8.689158(4)$\,s for the \nustar\ data without the need for demodulation (Figure~\ref{fig:z4_makishima}). This is consistent with the $P=8.689163(5)$\,s expected from the \emph{RXTE} ephemeris of \axp \citep[Ephemeris E, Table 6]{dib2014}. From the 2009 August 12--14, \emph{Suzaku}-XIS data, ME14 reported a period of $P=8.68891\pm0.00010$\,s, which is inconsistent with the value of $P=8.68869734(8)$\,s reported from the \emph{RXTE} ephemeris \citep[Ephemeris D, Table 6]{dib2014}. The \emph{Suzaku}-XIS and HXD measured rotation periods are marked in Figure~\ref{fig:z4_makishima} with their corresponding errors along with the \emph{RXTE} ephemeris for comparison.

Figure~\ref{fig:z4_makishima} indicates that in the \nustar\ observations, the pulsations were detected at a higher significance than during the high-energy \emph{Suzaku} observations. This result is (a) similar in value and shape to the result reported in Figure 1a of ME14 (XIS data from the same \emph{Suzaku} observations), (b) significantly higher than the $Z^2_3=12$ and $Z^2_4=16$ (without demodulation) and $Z^2_4\approx52$ (after optimal demodulation) reported in their Figures 1b and 1c. 

We searched for phase modulation in the data by shifting the arrival times of each photon by $\Delta t = A \sin(2\pi t/T-\phi_0)$, where $t$ is the time of arrival, $A$ is the modulation amplitude (with units of time), $T$ is the modulation period and $\phi_0$ is the initial phase. We measured $Z^2_4$ after varying $T$ between 45\,ks to 65\,ks in steps of 2.5\,ks, $A$ between 0--1.2\,s in steps of 0.1\,s  and $\phi_0$ between 0$^{\circ}$--360$^{\circ}$ in steps of 20$^{\circ}$. These results were compared to Figure 2 of ME14. We find that unlike ME14, $Z^2_4$ peaks to a value $>$350 at $A=0$ reducing to $\approx$100 at $A=1.2$.  $Z^2_4$ is also nearly independent of $\phi_0$ at any given $T$ and $A$. We find no preference for the values of $A=0.7\pm0.3$\,s and $\phi_0=75^{\circ}\pm30^{\circ}$.
 
Assuming a 55-ks period reported by ME14, we split the data into six subsets, each 9.17\,ks long, and created individual pulse profiles by folding each subset at $P=8.689158(4)$\,s. We find no phase change between any two pulse profiles (compared to Figure 3 of ME14). We also find that the post-demodulation pulse profile reported in Figure 1f of ME14 is triple-peaked with $PF_\mathrm{area}\lesssim10$\% and significantly different from the double-peaked \nustar\ profile and the 20--50\,keV pulse profile reported from \xmm\ data in dH08 (Figure 7E), each with $PF_\mathrm{area}\approx30$\%.

\subsection{Spectral Fits}
\label{sec:spectra_phase_average}
The phase-averaged X-ray spectrum of \axp\ has been previously fit with a hard power law at high energies ($\gtrsim$20\,keV) and by a modified blackblody (BB) or combination of blackbodies at low energies ($\lesssim$10\,keV). We fit the extracted spectrum in \texttt{XSPEC} with three different models: I) a hard high-energy power law (PL) plus a blackbody and a soft low-energy PL, II) a hard high-energy PL plus two BB models at low energies,
and III) a hard high-energy PL plus a comptonized BB \citep[\texttt{nthcomp}][]{zycki1999} at low energies. Each model included a \texttt{tbabs} model \citep{wilms2000} with solar elemental abundances (\texttt{`aspl'}) and cross-sections described by the \texttt{`bcmc'} model \citep{balucinskachurch1992,yan1998} to fit for photo-electric absorption and a cross-normalization parameter to allow for slight calibration differences between the \swift-XRT, \nustar\ FPMA and \nustar\ FPMB detectors. In Section~\ref{sec:model_fits}, we present fits to the phase-averaged spectra with a customized combination of model I and model II: a sum of one blackbody, one modified blackbody with a soft power-law tail and a hard power law\revision{, similar to the resonant compton scattering model used by \citet{rea2007}}.

The results of the fitting are shown in Table~\ref{tab:spectra_phase_average}. We checked the validity of each model fit by generating 1000 sets of synthetic data based on the best-fit model parameters and testing their $\chi^2$ with respect to the model. If the distribution of $\chi^2$ values from synthetic data is significantly lower than the $\chi^2$ of the real data, the fit is deemed to be unacceptable. The `goodness' parameter in Table~\ref{tab:spectra_phase_average} shows the fraction of synthetic $\chi^2$ values that are lower than the $\chi^2$ from the real data. Model I and II have traditionally been used to describe the spectrum of \axp. However, we find that while the models can match the spectral distribution visually, the fits are statistically unacceptable. Model III provides a statistically acceptable fit. 

\begin{figure}
\centering
\includegraphics[width=0.48\textwidth]{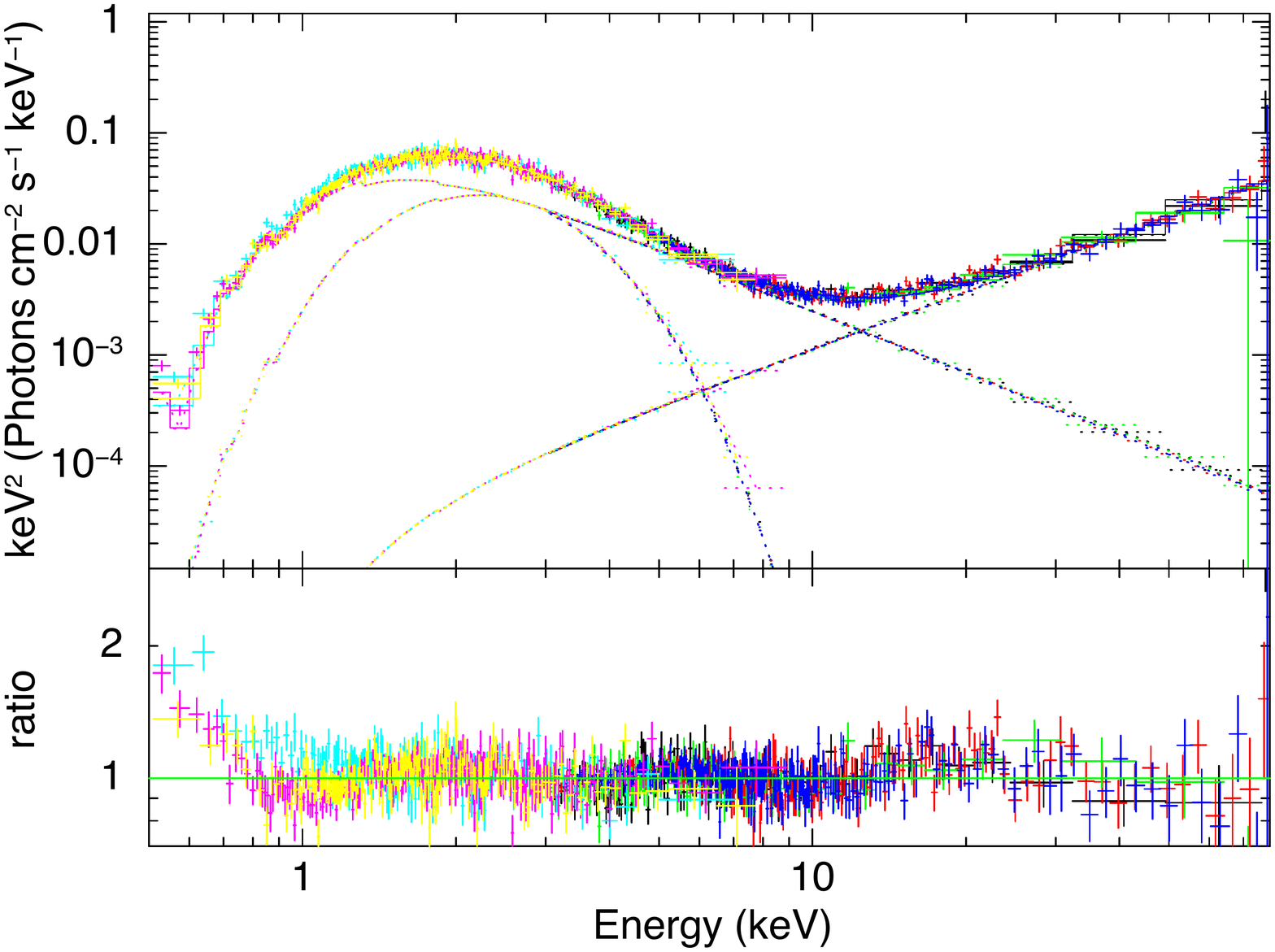}
\includegraphics[width=0.48\textwidth]{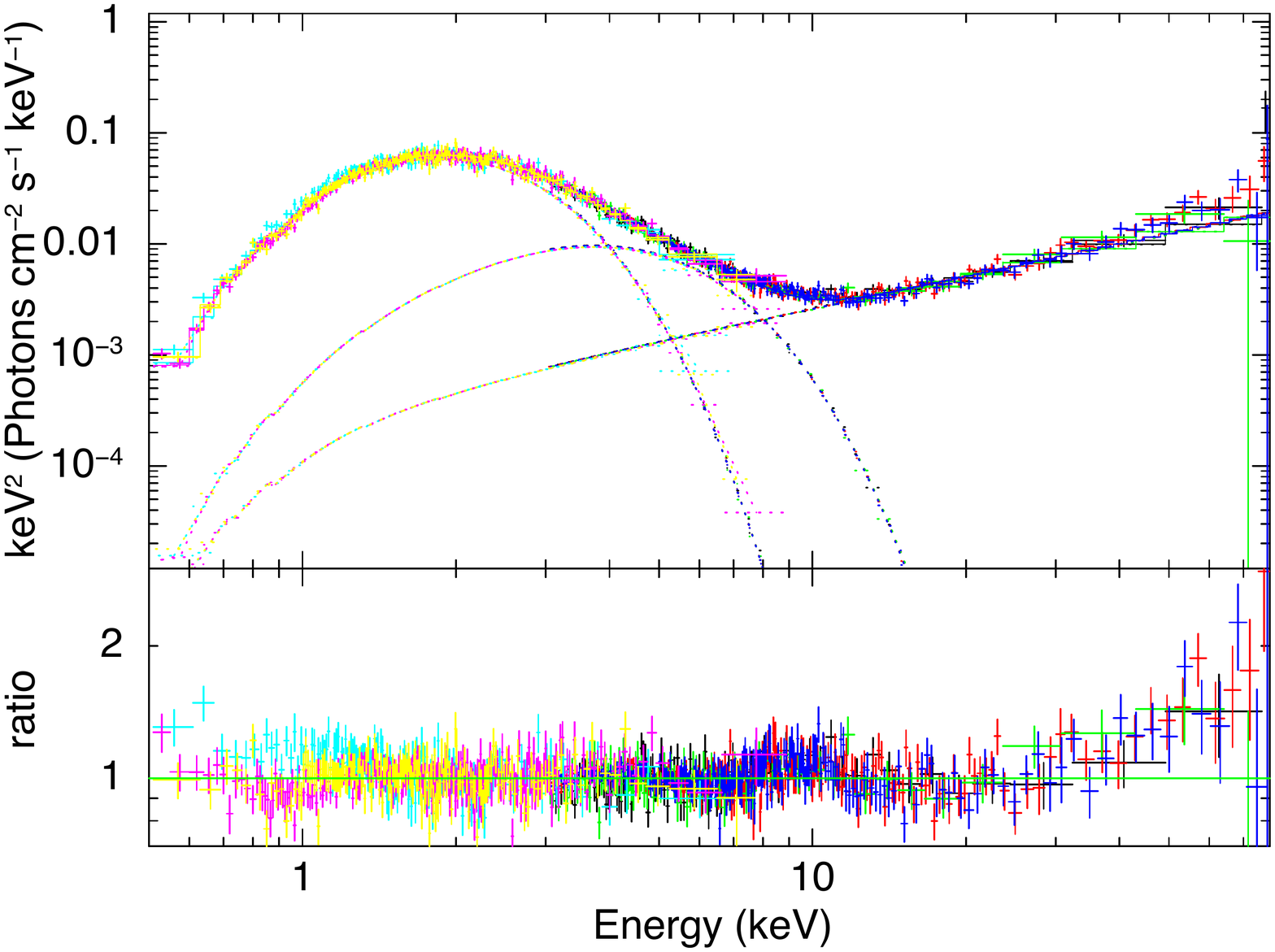}
\includegraphics[width=0.48\textwidth]{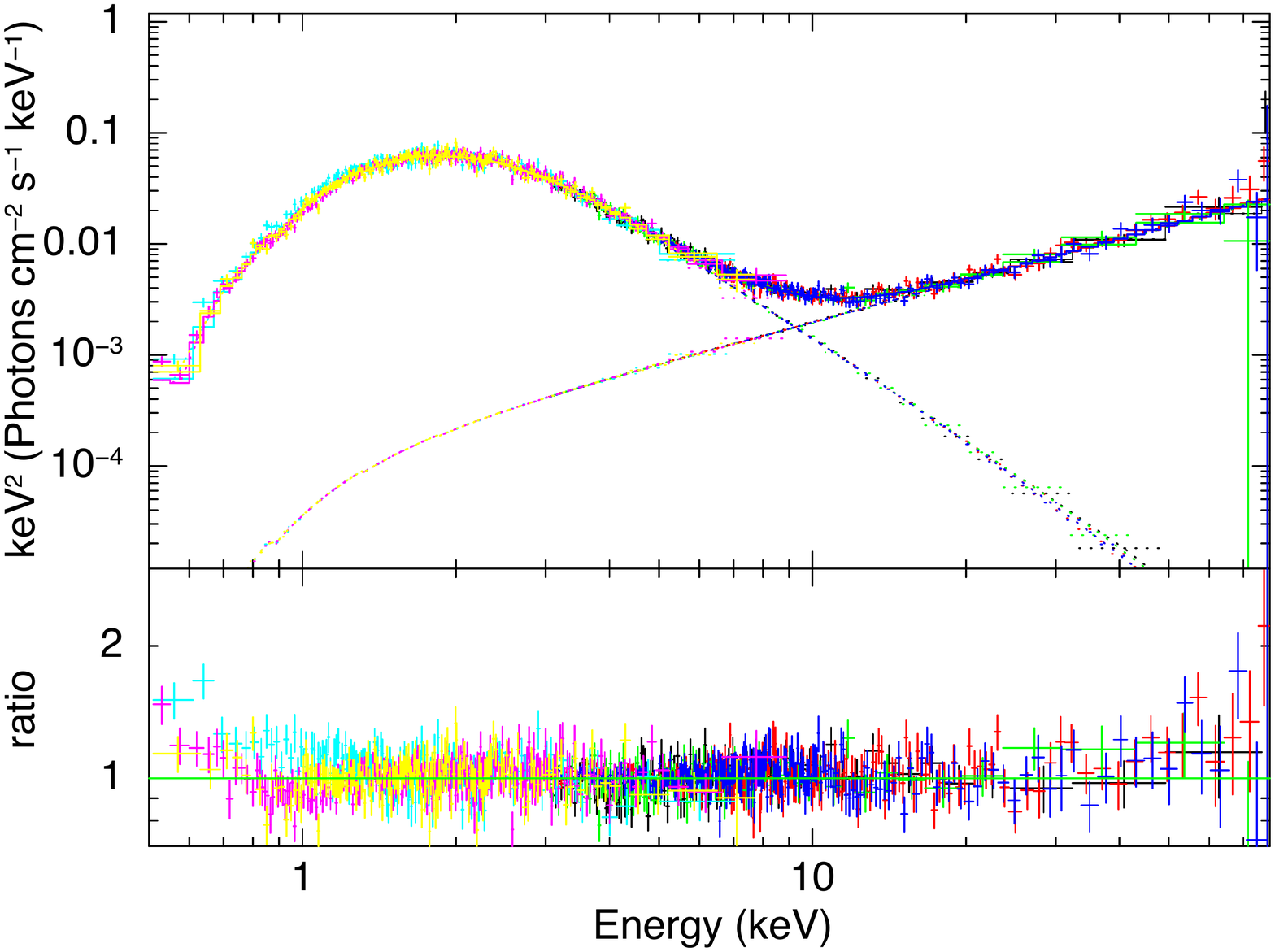}
\caption{Unfolded phase-averaged \swift-XRT and \nustar\ spectrum and the ratio of the data to the model. The model fit shown is \texttt{const*tbabs*(bbody+powerlaw+powerlaw)} (Model I, top panel), \texttt{const*tbabs*(bbody+bbody+powerlaw)} (Model II, middle panel) and \texttt{const*tbabs*(nthcomp+powerlaw)} (Model III, bottom panel). The colors are as follows: black: \nustar\ FPMA Obs I, red: \nustar\ FPMA Obs II, blue: \nustar\ FPMB Obs I, green: \nustar\ FPMB Obs II, cyan, yellow, magenta: \swift-XRT Obs I, II and III.}
\label{fig:const_tbabs_bbody_bbody}
\end{figure}

Figure~\ref{fig:const_tbabs_bbody_bbody} shows the fit of model I (BB+2PL, top panel), model II (2BB+PL, middle panel) and model III (\texttt{nthcomp}+PL, bottom panel). As noted in Table~\ref{tab:spectra_phase_average}, the index of the high-energy power law ($\Gamma_H$) varies between 0.3--1.0 depending on the model used to fit the low energy spectrum. This is reflected in the residuals at the high-energy end of Figure~\ref{fig:const_tbabs_bbody_bbody}. It is clear that the best-fit high-energy power law underpredicts the data at high energy for model II (middle panel) while it over predicts the data when the $<10\,$keV spectrum is modeled with model I (top panel). Note that while the \texttt{nthcomp} model is a good phenomenological fit to the low and intermediate energy X-ray spectrum and the blackbody emission from the surface is expected to be upscattered by high-energy electrons outside the neutron star, the \texttt{nthcomp} model does not accurately account for the effects of the extremely strong magnetic field on the photon scattering process.

If we restrict the fits to the low-energy spectrum ($<10$\,keV), model I fits improve with $\chi^2/\mathrm{dof}=1554.3/1391$ and the parameter values are similar (within 2-$\sigma$) to those in Table~\ref{tab:spectra_phase_average}, suggesting that this model can fit the low-energy spectrum well but cannot describe the 10--20\,keV region of the spectrum. Fitting model II to the low-energy spectrum produces $\chi^2/\mathrm{dof}=2082.4/1391$, which is statistically unacceptable.

Assuming a nominal value for the neutron star radius $R_\mathrm{NS}=10$\,km and a distance of 3.6\,kpc \citep{durant2006a}, we can calculate the fraction of the neutron star surface area ($\mathcal{A}_{\rm NS}$) covered by the blackbody of a given flux normalization. For model I, the blackbody has a bolometric luminosity of $L_\mathrm{bol}=1.3\times10^{35}\,\mathrm{erg\,s^{-1}}$, covering $0.2\,\mathcal{A}_{\rm NS}$ and contributing 11\% of the 0.5--79\,keV X-ray luminosity (most of it in the 0.5--10\,keV band), while the soft power law contributes the remaining 84\%. For model II, the low temperature blackbody ($L_\mathrm{bol}=2.5\times10^{35}\,\mathrm{erg\,s^{-1}}$) covers $0.6\,\mathcal{A}_{\rm NS}$ and contributes 75\% of the luminosity, while the high temperature blackbody has a luminosity of $L_\mathrm{bol}=3.3\times10^{34}\,\mathrm{erg\,s^{-1}}$ emanating from a hotspot covering $0.004\,\mathcal{A}_{\rm NS}$ of the surface and contributing 10\% of the X-ray luminosity. In model III, the \texttt{nthcomp} component (combined blackbody and comptonized power law) contribute 86\% of the total X-ray flux, similar to the contributions of the two blackbodies of model II.

\begin{deluxetable}{llc}
\tablecolumns{3}
\tablecaption{Phase-averaged spectral fits. \label{tab:spectra_phase_average}}
\tablewidth{0pt}
\tabletypesize{\footnotesize}
\tablehead{
  \colhead{Component} &
  \colhead{Parameter}   &
  \colhead{Value} 
}
\startdata
\sidehead{Model I}
\sidehead{\texttt{const*tbabs*(bbody+powerlaw+powerlaw)}}
\texttt{const}    & $C_{\mathrm{FPMA}}$ & $0.981\pm0.015$  \\
                  & $C_{\mathrm{FPMB}}$ & $0.977\pm0.015$  \\
\texttt{tbabs}    & $N_\mathrm{H}$ ($10^{22}\,\mathrm{cm^{-2}}$) & $1.30\pm0.03$ \\
\texttt{bbody}    & $k_BT_{\mathrm{BB}}$ (keV) & $0.462\pm0.005$ \\
                  & norm\tablenotemark{a} ($10^{-3}$) & $1.06\pm0.04$ \\
                  & $F_\mathrm{BB}$\tablenotemark{c} & 0.11 \\ 
\texttt{powerlaw} & $\Gamma_S$ & $3.85\pm0.04$  \\
                  & norm\tablenotemark{b} & $0.18\pm0.01$ \\
                  & $F_\mathrm{PL,S}$\tablenotemark{c} & 0.84 \\ 
\texttt{powerlaw} & $\Gamma_H$ & $0.29\pm0.05$  \\
                  & norm\tablenotemark{b} ($10^{-5}$) & $2.3\pm0.4$ \\
                  & $F_\mathrm{PL,H}$\tablenotemark{c} & 0.05 \\ 
                  \texttt{$\chi^2_{\mathrm{red}}/\mathrm{dof}$} & & 1.174/2408 \\
                  $p$-value & & $4.8\times10^{-9}$ \\
                  goodness\tablenotemark{d} & & 100\% \\
\hline
\sidehead{Model II}
\sidehead{\texttt{const*tbabs*(bbody+bbody+powerlaw)}}
\texttt{const}    & $C_{\mathrm{FPMA}}$ & $1.033\pm0.016$  \\
                  & $C_{\mathrm{FPMB}}$ & $1.029\pm0.016$  \\
\texttt{tbabs}    & $N_\mathrm{H}$ ($10^{22}\,\mathrm{cm^{-2}}$) & $0.52\pm0.01$  \\
\texttt{bbody}    & $k_BT_{\mathrm{BB,1}}$ (keV) & $0.422\pm0.004$                 \\
                  & norm\tablenotemark{a} ($10^{-3}$) & $1.90\pm0.02$      \\
                  & $F_\mathrm{BB,1}$\tablenotemark{c} & 0.75 \\ 
\texttt{bbody}    & $k_BT_{\mathrm{BB,2}}$ (keV) & $0.93\pm0.02$                   \\
                  & norm\tablenotemark{a} ($10^{-4}$) & $2.5\pm0.1$        \\
                  & $F_\mathrm{BB,2}$\tablenotemark{c} & 0.10 \\ 
\texttt{powerlaw} & $\Gamma_H$ & $1.03\pm0.05$                                   \\
                  & norm\tablenotemark{b} ($10^{-4}$) & $2.7_{-0.3}^{+0.4}$  \\
                  & $F_\mathrm{PL,H}$\tablenotemark{c} & 0.15 \\ 
                  \texttt{$\chi^2_{\mathrm{red}}/\mathrm{dof}$} & & 1.130/2408     \\
                  $p$-value & & $6.7\times10^{-6}$ \\
                  goodness\tablenotemark{d} & & 99.7\% \\
\hline
\sidehead{Model III}
\sidehead{\texttt{const*tbabs*(nthcomp+powerlaw)}}
\texttt{const}    & $C_{\mathrm{FPMA}}$ & $1.001\pm0.015$  \\
                  & $C_{\mathrm{FPMB}}$ & $0.998\pm0.015$  \\
\texttt{tbabs}    & $N_\mathrm{H}$ ($10^{22}\,\mathrm{cm^{-2}}$) & $0.65\pm0.02$  \\
\texttt{nthcomp}  & $\Gamma_S$ & $4.86\pm0.04$                             \\
                  & $k_BT_{\mathrm{BB}}$ (keV) &  $ 0.346\pm0.004$           \\
                  & $k_BT_{e^{-}}$ (keV) &  $>37.3$                          \\
                  & norm\tablenotemark{b} ($10^{-2}$) & $6.5\pm0.2$  \\
                  & $F_\mathrm{nthcomp}$\tablenotemark{c} & 0.86 \\ 
\texttt{powerlaw} & $\Gamma_H$ & $0.75_{-0.04}^{+0.05}$                            \\
                  & norm\tablenotemark{b} ($10^{-4}$) & $1.1_{-0.1}^{+0.2}$ \\
                  & $F_\mathrm{PL}$\tablenotemark{c} & 0.14 \\ 
                  \texttt{$\chi^2_{\mathrm{red}}/\mathrm{dof}$} & & 1.07/2408  \\
                  $p$-value & & $6.3\times10^{-3}$ \\
                  goodness\tablenotemark{d} &  & 82.8\%         
\enddata
\tablenotetext{a}{Normalization in units of $L_{39}/D_{10}^2$, where $L_{39}$ is the source luminosity in units of $10^{39}\,\mathrm{erg\,s^{-1}}$ and $D_{10}$ is the distance to the source in units of 10\,kpc.}
\tablenotetext{b}{Normalization in units of $\mathrm{photons\,keV^{-1}\,cm^{-2}\,s^{-1}}$ at 1\,keV.}
\tablenotetext{c}{Fraction of the total 0.5--79\,keV flux contributed by the component.}
\tablenotetext{d}{Goodness of fit is the percentage of $\chi^2$ values from 1000 Monte Carlo simulations synthesized the best-fit model parameters that are less than the best-fit $\chi^2$ value. The data are indistinguishable from the synthesized data if the goodness $\approx$50\%.}
\end{deluxetable}

\subsubsection{High-Energy Power Law}
\label{sec:highenergypowerlaw}
The differences in the hard power law between different models are caused by the inability of these phenomenological models to accurately describe the spectrum between approximately 10--20\,keV. In order to minimize the figure-of-merit ($\chi^2$ in this case) for the fit, \texttt{XSPEC} forces variations in the hard power-law index and normalization. To better measure the slope of the hard power law, we restricted the energy range from 20--79\,keV and fit the phase-averaged spectrum with a power law. We measure $\Gamma_H=0.65\pm0.09$. The 2 parameter fit yielded a $\chi^2_\mathrm{red}=477.12$ for 492 degrees of freedom and a $p$-value of 0.68. This is independent of $N_\mathrm{H}$ and the model used to describe the soft X-ray ($<10$\,keV) spectrum. When the energy range is further constrained (i.e. in the ranges 25--79\,keV and 30--79\,keV), we get consistent measures of $\Gamma_H$ but with larger uncertainties. This value is lower than the $\Gamma_H=0.93\pm0.06$ measured by dH08 and $\Gamma_H=0.89\pm0.10$ measured by \citet{enoto2011}. However, note that the values measured by dH08 varied from $0.79\pm0.10$ to $1.21\pm0.16$ over different datasets. The \texttt{nthcomp} model provides the least structured residuals and the value of $\Gamma_H$ is closest to the high-energy-only value. 

If the high-energy hard power law is frozen to the $\Gamma_H=0.65$ value and the corresponding normalization and the low-energy (0.5--10\,keV) spectrum is described with parameters from model I, the fit worsens with $\chi^2=3104.0$ in 2412 degrees of freedom. The fit for model II also worsens with $\chi^2=3061.3$ in 2412 degrees of freedom. Both models show structured wavy residuals between 5--11\,keV suggesting that the models are failing to capture all the structure in the data. Model III fit parameters do not change values within 1-$\sigma$ errors when the hard power law is frozen, $\chi^2=2602.8$ for 2412 degrees of freedom and the goodness of fit is 85\%.
    
\subsubsection{Photoelectric Absorption}
\label{sec:photoelectricabsorption}
We note that the  $N_\mathrm{H}$ value from Model III (Table~\ref{tab:spectra_phase_average}) is consistent with the D06b value of $(6.4\pm0.7)\times10^{21}\,\mathrm{cm^{-2}}$. To check on the influence of various abundance models, we re-tested the model fits with the seven abundance data sets used in X-ray astronomy (Table~\ref{tab:abundance_models}, in chronological order). We find that the older abundance data sets (\texttt{aneb}, \texttt{angr}) consistently provide lower $N_\mathrm{H}$ values compared to the newer data sets (\texttt{wilm},\texttt{lodd}, \texttt{aspl}) and that the \texttt{nthcomp}+PL model provides the better fit irrespective of which abundance model is used. Note that these values are roughly similar to the previous values reported by \citet{patel2003} ($N_\mathrm{H}=(0.93\pm0.02)\times10^{22}\,\mathrm{cm^{-2}}$, using \texttt{aneb} abundances and Model I) and by \citet{rea2007} ($N_\mathrm{H}=(0.926\pm0.005)\times10^{22}\,\mathrm{cm^{-2}}$, using \texttt{angr} abundances and Model I).

\begin{deluxetable*}{lccccccccc}
\tablecolumns{10}
\tablecaption{$N_\mathrm{H}$ values from different abundance models. \label{tab:abundance_models}}
\tablewidth{0pt}
\tabletypesize{\footnotesize}
\tablehead{
  \colhead{Abund.\tablenotemark{a}} &
  \multicolumn{3}{c}{Model I} & 
  \multicolumn{3}{c}{Model II} & 
  \multicolumn{3}{c}{Model III} \\
  \colhead{Model} &
  \colhead{$N_\mathrm{H}$\tablenotemark{b}} &
  \colhead{$\chi^2$\tablenotemark{c}} &
  \colhead{$p$-value} &
  \colhead{$N_\mathrm{H}$\tablenotemark{b}} &
  \colhead{$\chi^2$\tablenotemark{c}}&
  \colhead{$p$-value} &
  \colhead{$N_\mathrm{H}$\tablenotemark{b}} &
  \colhead{$\chi^2$\tablenotemark{c}} &
  \colhead{$p$-value} 
}
\startdata
\texttt{aneb} & $1.06\pm0.02$ & 2725.33 & $5.5\times10^{-6}$ & $0.41\pm0.01$ & 2722.70 & $6.5\times10^{-6}$ & $0.52\pm0.01$ & 2553.53 & $1.9\times10^{-2}$ \\
\texttt{angr} & $0.93\pm0.02$ & 2719.04 & $8.1\times10^{-6}$ & $0.37\pm0.01$ & 2718.96 & $8.1\times10^{-6}$ & $0.46\pm0.01$ & 2551.61 & $2.1\times10^{-2}$ \\
\texttt{feld} & $0.95\pm0.02$ & 2768.78 & $3.7\times10^{-7}$ & $0.38\pm0.01$ & 2719.98 & $7.6\times10^{-6}$ & $0.47\pm0.01$ & 2567.20 & $1.2\times10^{-2}$ \\
\texttt{grsa} & $1.11\pm0.02$ & 2753.07 & $9.6\times10^{-7}$ & $0.44\pm0.01$ & 2719.87 & $7.7\times10^{-6}$ & $0.55\pm0.01$ & 2563.12 & $1.4\times10^{-2}$ \\
\texttt{wilm} & $1.27\pm0.03$ & 2826.47 & $5.3\times10^{-9}$ & $0.51\pm0.01$ & 2722.94 & $6.4\times10^{-6}$ & $0.64\pm0.02$ & 2588.34 & $5.4\times10^{-3}$ \\
\texttt{lodd} & $1.32\pm0.03$ & 2843.74 & $1.4\times10^{-9}$ & $0.54\pm0.01$ & 2718.83 & $8.2\times10^{-6}$ & $0.67\pm0.02$ & 2592.26 & $4.7\times10^{-3}$ \\
\texttt{aspl} & $1.30\pm0.03$ & 2827.83 & $4.8\times10^{-9}$ & $0.52\pm0.01$ & 2722.14 & $6.7\times10^{-6}$ & $0.65\pm0.02$ & 2584.44 & $6.3\times10^{-3}$ 
\enddata
\tablenotetext{a}{References --- \texttt{aneb}: \citet{anders1982}, \texttt{angr}: \citet{anders1989}, \texttt{feld}: \citet{feldman1992}, \texttt{grsa}: \citet{grevesse1998}, \texttt{wilm}: \citet{wilms2000}, \texttt{lodd}: \citet{lodders2003}, \texttt{aspl}: \citet{asplund2009}.}
\tablenotetext{b}{In units of $10^{22}\,\mathrm{cm^{-2}}$.}
\tablenotetext{c}{Each fit has 2408 degrees of freedom.}
\end{deluxetable*}

\subsubsection{Freezing $N_\mathrm{H}$ and the High-Energy PL}
The D06b value of $N_\mathrm{H}$ and our measurement of the high-energy PL are independent of the complicated spectral shape at low and intermediate energies ($<20$\,keV). By freezing the value of $N_\mathrm{H}=6.4\times10^{21}\,\mathrm{cm^{-2}}$ and freezing the high-energy power law to the slope and normalization measured in Section~\ref{sec:highenergypowerlaw}, we can explore the low-energy spectral shape and investigate whether additional spectral components are required to fully describe the low-energy distribution. For technical reasons\footnote{Since the $N_\mathrm{H}$ value affects only the \swift-XRT spectrum and the high-energy PL affects only the \nustar\ spectrum, allowing the cross-normalization factors to vary freely effectively allows the high-energy PL normalization to vary, spoiling the high-energy fit. To prevent this effect, we must freeze the cross-normalization constants. The expected systematic cross-calibration error between \nustar\ and \swift-XRT is approximately 5\% \citep{madsen2015}.}, the cross-normalization factors between \swift-XRT and \nustar\ were frozen to unity. 

\begin{figure}
\centering
\includegraphics[width=0.48\textwidth]{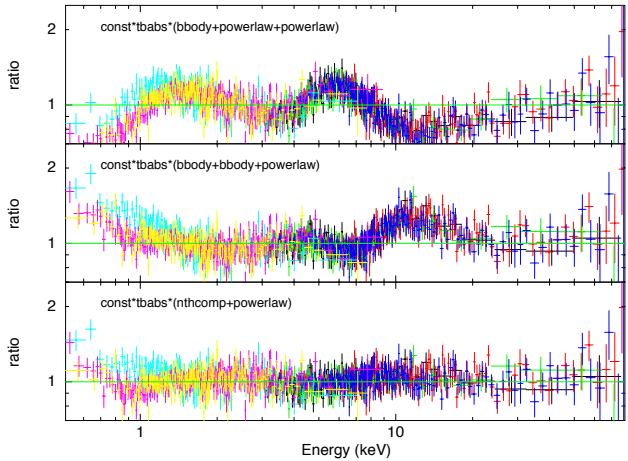}
\caption{Data to model ratio for models fit with $N_\mathrm{H}$ and hard power-law parameters frozen to independently measured values (see Section~\ref{sec:photoelectricabsorption}). From top to bottom, the plots represent model I, model II and model III.}
\label{fig:frozen_nh_residuals}
\end{figure}

We find that Model I and Model II fits worsen significantly with $\chi^2=5896.6$ and $\chi^2=3480.0$ for 2413 degrees of freedom respectively with extremely wavy residuals (Figure~\ref{fig:frozen_nh_residuals}, Panels I and II). Model III provides a fit parameters similar to that from Table~\ref{tab:spectra_phase_average} with a total $\chi^2=2608.3$ for 2413 degrees of freedom.



\subsection{Phase-Resolved Spectral Fits}
We created good-time-interval (\texttt{gti}) files using the measured period of \axp\ and extracted \swift-XRT and \nustar\ spectra in five equal phase bins: $\phi=$0.0--0.2, 0.2--0.4, 0.4--0.6, 0.6--0.8 and 0.8--1.0. We fit each 0.5--79\,keV spectrum with a BB+2PL and \texttt{nthcomp} + PL models. The fit parameters are detailed in Table~\ref{tab:spectra_phase_resolved}. We froze the values of $C_\mathrm{FPMA}$, $C_\mathrm{FPMB}$ and $N_\mathrm{H}$ to those fit in the phase-averaged spectrum using the same spectral model (as in Table~\ref{tab:spectra_phase_average}). 

\begin{figure*}
\centering
\includegraphics[clip=true,trim=0.2in 0.00in 0.6in 0.4in,width=0.33\textwidth]{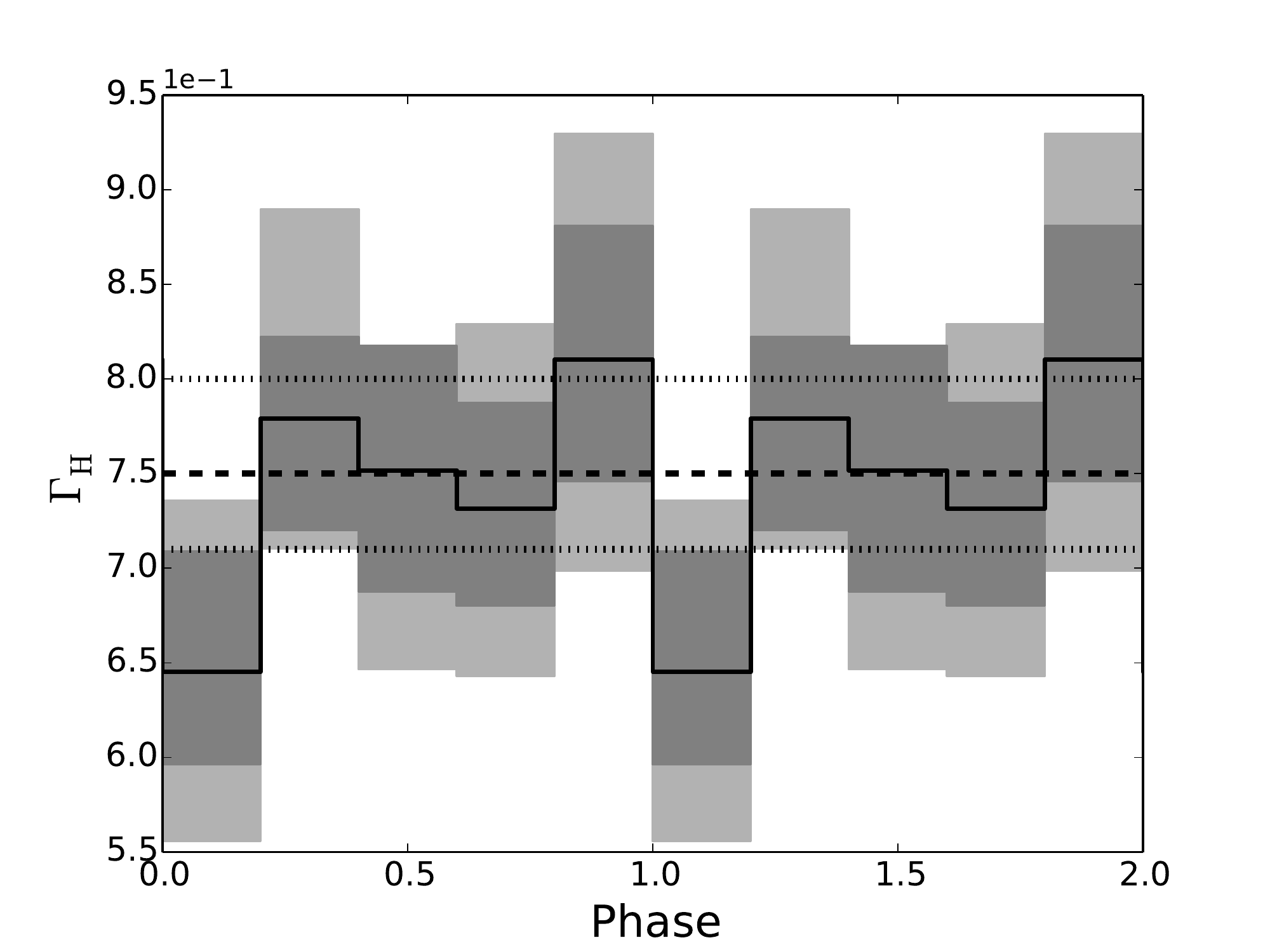}
\includegraphics[clip=true,trim=0.2in 0.00in 0.6in 0.4in,width=0.33\textwidth]{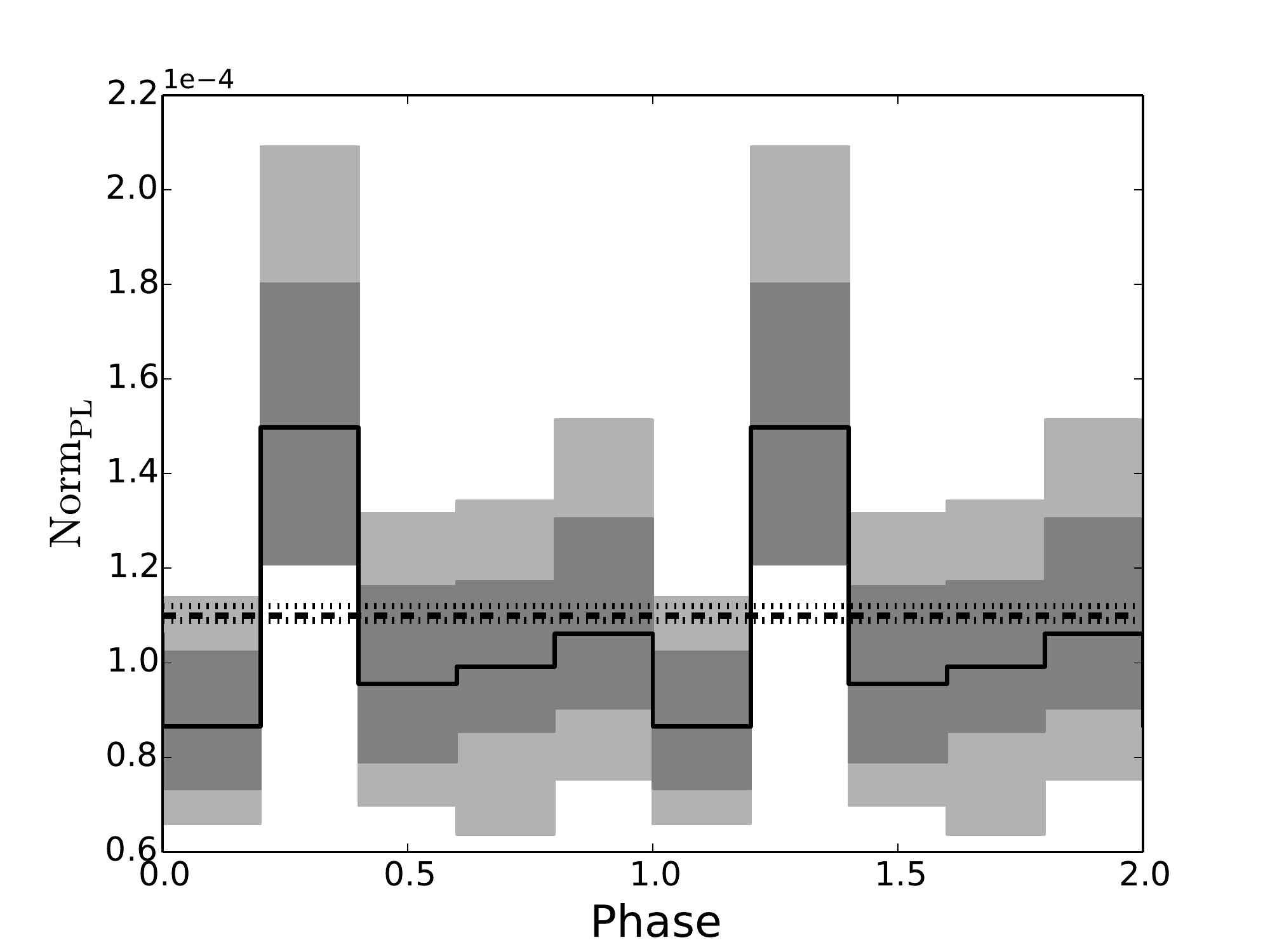}
\includegraphics[clip=true,trim=0.2in 0.00in 0.6in 0.4in,width=0.33\textwidth]{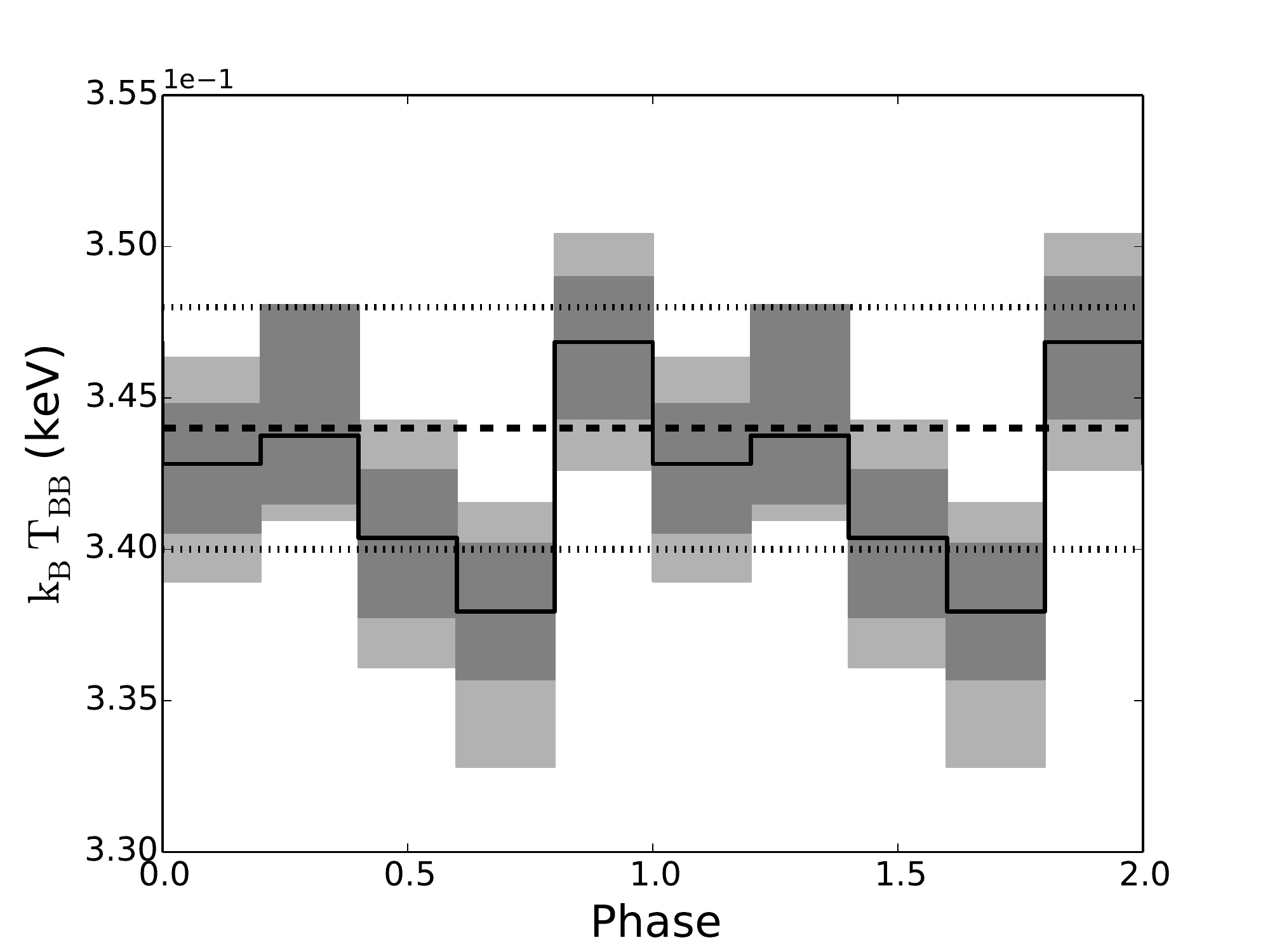}
\includegraphics[clip=true,trim=0.2in 0.00in 0.6in 0.4in,width=0.33\textwidth]{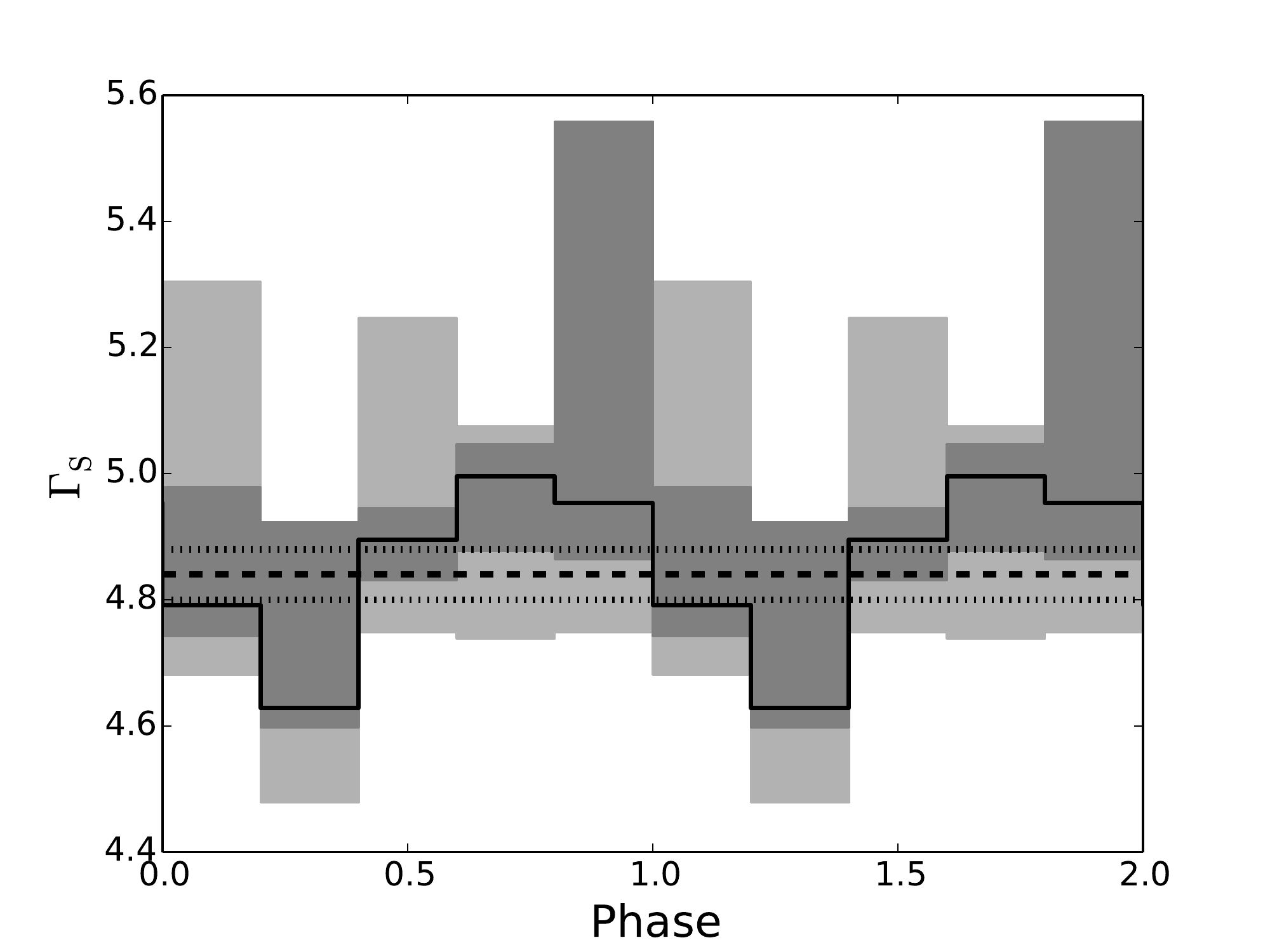}
\includegraphics[clip=true,trim=0.2in 0.00in 0.6in 0.4in,width=0.33\textwidth]{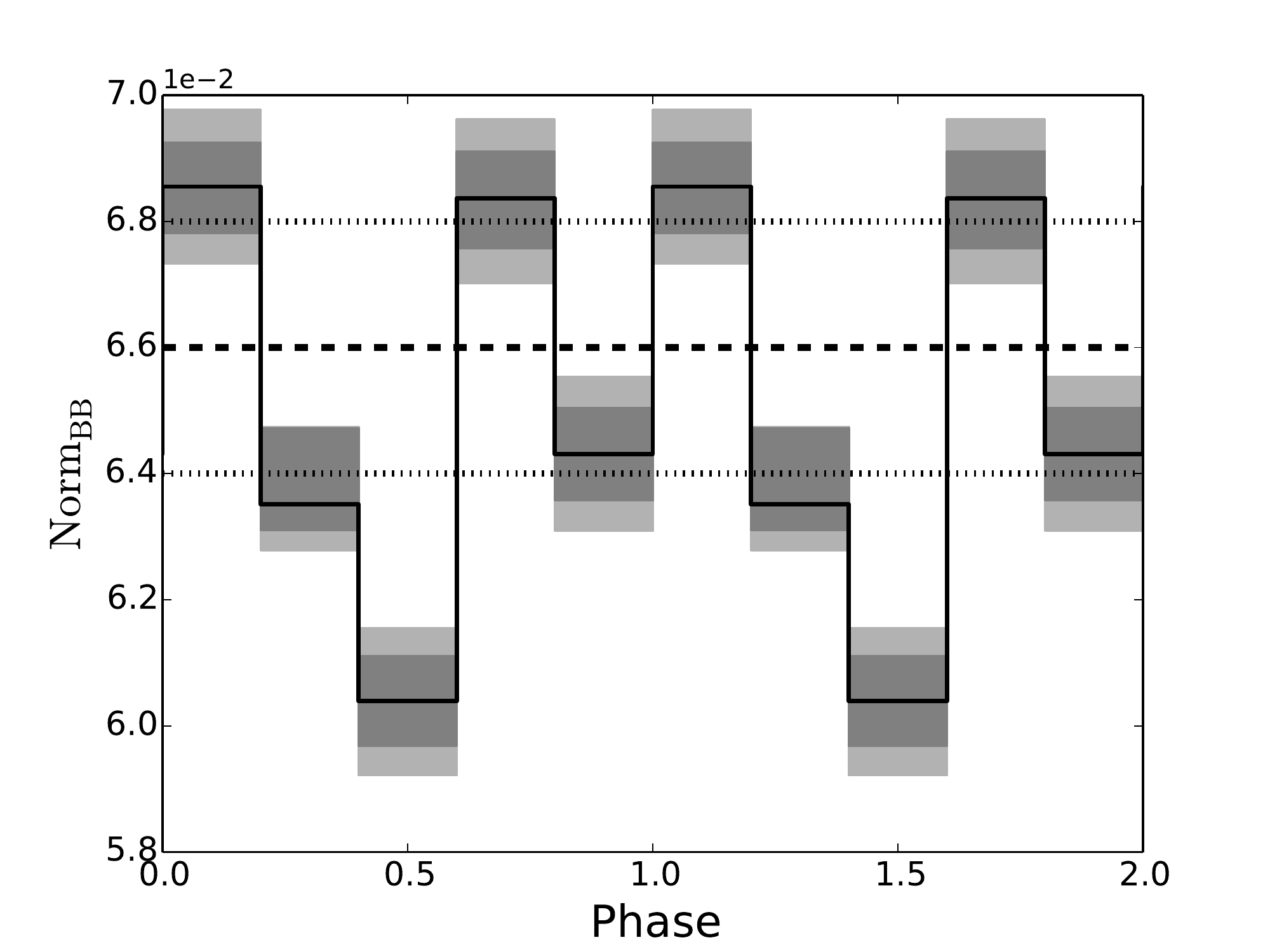}
\caption{Variation of \texttt{nthcomp}+PL model parameters as a function of rotational phase. In each plot, the solid black line shows the parameter value in each phase bin, dark and light gray regions show the 1-$\sigma$ and 3-$\sigma$ error ranges respectively. The phase range is repeated twice for clarity. The dashed black line and dotted black lines show the value of the parameter in the phase-averaged spectral fit and the corresponding 3-$\sigma$ error bars. Starting from the top left to bottom right, the plots show $\Gamma_H$, power law normalization, $k_BT_\mathrm{BB}$, $\Gamma_S$ and \texttt{nthcomp} normalization, respectively.}
\label{fig:spectra_phase_resolved}
\end{figure*}

The $\chi^2_\mathrm{red}$ for each individual phase is lower than that from the corresponding fits of the phase-averaged spectra. Figure~\ref{fig:spectra_phase_resolved} shows fit parameters for the \texttt{nthcomp}+PL model as a function of phase compared to the phase-averaged fit values. The spectral shape parameters $\Gamma_H$, $k_BT_\mathrm{BB}$ and $\Gamma_S$ are statistically consistent within 3-$\sigma$ with the values measured from the phase-averaged spectra. However, we detect a very significant increase in the hard power-law normalization in the 0.2--0.4 phase range which corresponds to the peak of the high-energy pulse profiles (20--35\,keV, 35--50\,keV, Figure~\ref{fig:pulse_profile}). Similarly, the normalization of the \texttt{nthcomp} component shows a sharp decrease in the 0.4--0.6 phase bin which corresponds to the dip at $\phi=0.5$ in the 3--5\,keV and 5--8\,keV pulse profiles. For the BB+2PL model, we observe a similar trend with the hard power-law normalization significantly increasing in the 0.2--0.4 phase range and the soft power-law normalization (which contributes approximately 85\% of the X-ray flux) decreasing between the 0.4--0.6 phase range. In Section~\ref{sec:model_fits}, we describe the variation in the high-energy spectra in greater detail with a physical emission model.

\begin{deluxetable*}{llcccccc}
\centering
\tablecolumns{8}
\tablecaption{Spectral fits to 0.5--79\,keV phase-resolved \swift-XRT and \nustar\ observations.\label{tab:spectra_phase_resolved}}
\tablewidth{0pt}
\tabletypesize{\footnotesize}
\tablehead{
  \colhead{Component} &
  \colhead{Parameter}   &
  \multicolumn{6}{c}{Phase Range}\\
  \colhead{} &
  \colhead{} &
  \colhead{0.0--1.0} &
  \colhead{0.0--0.2} &
  \colhead{0.2--0.4} &
  \colhead{0.4--0.6} &
  \colhead{0.6--0.8} &
  \colhead{0.8--1.0} 
}
\startdata
\sidehead{\texttt{const*tbabs*(bbody+powerlaw+powerlaw)}}
\texttt{const}    & $C_{\mathrm{FPMA}}$ & $0.981\pm0.015$ & -- & -- & -- & -- & -- \\
                  & $C_{\mathrm{FPMB}}$ & $0.977\pm0.015$ & -- & -- & -- & -- & -- \\
\texttt{tbabs}    & $N_\mathrm{H}$ ($10^{22}\,\mathrm{cm^{-2}}$) & $1.30\pm0.03$ & -- & -- & -- & -- & -- \\
\texttt{bbody}    & $k_BT_{\mathrm{BB}}$ (keV)                 & $0.462\pm0.005$  & $0.474_{-0.008}^{+0.008}$ & $0.483_{-0.009}^{+0.009}$ & $0.467_{-0.008}^{+0.009}$ &  $0.458_{-0.008}^{+0.008}$ & $0.478_{-0.008}^{+0.008}$ \\
                  & norm\tablenotemark{a} ($10^{-3}$) & $1.06\pm0.04$   & $ 0.90_{-0.06}^{+0.06} $ & $ 1.05_{-0.06}^{+0.07} $ & $ 0.988_{-0.06}^{+0.06} $& $ 1.02_{-0.06}^{+0.06} $ & $ 0.90_{-0.06}^{+0.06} $ \\
\texttt{powerlaw} & $\Gamma_S$                              & $3.85\pm0.03$  & $3.86_{-0.03}^{+0.03}$ & $3.84_{-0.03}^{+0.03}$ & $3.91_{-0.03}^{+0.03}$ & $3.97_{-0.03}^{+0.03}$ & $3.92_{-0.03}^{+0.03}$ \\
                  & norm\tablenotemark{b}                  & $0.18\pm0.01$  & $0.202_{-0.007}^{+0.006}$ & $0.192_{-0.006}^{+0.006}$ & $0.179_{-0.006}^{+0.006}$ & $0.200_{-0.007}^{+0.007}$ & $0.193_{-0.006}^{+0.006}$\\
\texttt{powerlaw} & $\Gamma_H$                             & $0.29\pm0.05$   & $0.2_{-0.1}^{+0.1}$ & $0.4_{-0.1}^{+0.1}$ & $0.3_{-0.1}^{+0.1}$ & $0.3_{-0.1}^{+0.1}$ &  $0.3_{-0.1}^{+0.1}$\\
                  & norm\tablenotemark{b} ($10^{-5}$) & $2.3\pm0.4$  & $2.1_{-0.6}^{+0.8}$ & $4.6_{-1.2}^{+1.6}$ & $2.2_{-0.7}^{+1.0}$ & $2.6_{-0.7}^{+1.0}$ & $2.1_{-0.7}^{+1.1}$\\
                  $\chi^2/\mathrm{dof}$ & & 2718.6/2408 & 1184.8/1122 & 1207.4/1120 & 1011.0/1008 & 1154.7/1045 & 1070.2/1035  \\
                  $p$-value & & $4.8\times10^{-9}$ & $9.4\times10^{-2}$ & $3.5\times10^{-2}$ & $4.7\times10^{-1}$ & $9.8\times10^{-3}$ & $2.2\times10^{-1}$   \\          
\hline
\sidehead{\texttt{const*tbabs*(nthcomp+powerlaw)}}
\texttt{const}    & $C_{\mathrm{FPMA}}$ & $1.001\pm0.015$ & -- & -- & -- & -- & -- \\
                  & $C_{\mathrm{FPMB}}$ & $0.998\pm0.015$ & -- & -- & -- & -- & -- \\
\texttt{tbabs}    & $N_\mathrm{H}$ ($10^{22}\,\mathrm{cm^{-2}}$) & $0.65\pm0.02$ & -- & -- & -- & -- & -- \\
\texttt{nthcomp}  & $\Gamma_S$               & $4.86\pm0.04$                        & $4.81_{-0.13}^{+0.06} $ & $4.75_{-0.30}^{+0.07} $ & $4.89_{-0.15}^{+0.08} $& $4.98_{-0.22}^{+0.08} $ & $4.94_{-0.21}^{+0.10} $\\
                  & $k_BT_{\mathrm{BB}}$ (keV) &  $ 0.346\pm0.004$                   & $0.344_{-0.009}^{+0.003}$ & $0.344_{-0.006}^{+0.004}$ & $0.340_{-0.004}^{+0.004}$ & $0.338_{-0.004}^{+0.004}$ & $0.346_{-0.004}^{+0.003}$ \\
                  & $k_BT_{e^{-}}$ (keV)       &  $>37.3$                            & $>13.5$ & $>4.8 $ & $>10.5$ &   $>6.7 $ & $>6.8 $ \\
                  & norm\tablenotemark{b} ($10^{-2}$) & $6.5\pm0.2$          & $6.8\pm0.1$ & $6.4\pm0.1$ & $6.0\pm0.1$ & $6.8\pm0.1$ & $6.4\pm0.1$ \\
\texttt{powerlaw} & $\Gamma_H$ & $0.75_{-0.04}^{+0.05}$                             & $0.65_{-0.1}^{+0.09}$ & $0.79_{-0.08}^{+0.1}$ & $0.75_{-0.11}^{+0.11}$ & $0.71_{-0.09}^{+0.1}$ & $0.78_{-0.09}^{+0.2}$ \\
                  & norm\tablenotemark{b} ($10^{-4}$) & $1.1_{-0.1}^{+0.2}$   &$ 0.9_{-0.2}^{+0.3} $ &$ 1.4_{-0.3}^{+0.6} $ &$ 1.0_{-0.3}^{+0.3} $& $ 0.9_{-0.2}^{+0.3} $ &$ 1.0_{-0.2}^{+0.6} $ \\
                  $\chi^2/\mathrm{dof}$ & & 2550.1/2408  & 1190.4/1122 & 1189.4/1120 & 1012.0/1008 & 1094.1/1045 & 1058.8/1035 \\
                  $p$-value & & $6.3\times10^{-3}$ & $7.6\times10^{-2}$ & $7.3\times10^{-2}$ & $4.6\times10^{-1}$ & $1.4\times10^{-1}$ & $3.0\times10^{-1}$ 
\enddata
\tablenotetext{a}{Normalization in units of $L_{39}/D_{10}^2$, where $L_{39}$ is the source luminosity in units of $10^{39}\,\mathrm{erg\,s^{-1}}$ and $D_{10}$ is the distance to the source in units of 10\,kpc.}
\tablenotetext{b}{Normalization in units of $\mathrm{photons\,keV^{-1}\,cm^{-2}\,s^{-1}}$ at 1\,keV.}
\end{deluxetable*}

\subsection{\AB{$e^\pm$ Outflow Model}}
\label{sec:model_fits}
Next we test the coronal outflow model proposed by \citet{beloborodov2013b}.
\AB{The}
model envisions an outflow of relativistic electron-positron ($e^{\pm}$) pairs created by 
\AB{electric}
discharge near the neutron star. The outflow moves along the magnetic field lines and gradually decelerates as it (resonantly) scatters the thermal X-rays. The outflow fills the active ``$j$-bundle''
\AB{that carries the electric currents of twisted}
magnetospheric field lines \citep{beloborodov2009}. It radiates most of its kinetic energy in hard X-rays before the $e^{\pm}$ pairs reach the top of the 
\AB{twisted}
magnetic loop and annihilate.

The magnetic dipole moment of \fouru\ is $\mu \approx 1.3 \times 10^{32} \ {\rm G\,cm}^{3}$ \citep[calculated from the spin-down rate;][]{dib2014}. Similar to \citet{hascoet2014}, we assume a simple geometry where the $j$-bundle is axisymmetric around the magnetic dipole axis. However, instead of assuming that the $j$-bundle emerges from a polar cap, its footprint is allowed to have a ring shape. \revision{The assumption of axisymmetry reduces the number of free parameters and appears to be sufficient to fit the phase-resolved spectra. In future work, the energy-resolved pulse profiles could be included in the fit to constrain the axial distribution of the $j$-bundle.}

\AB{This more general}
model has the following parameters:
(1) the power $\lj$ of the $e^{\pm}$ outflow along the $j$-bundle,
(2) the angle $\alphamag$ between the rotation axis and the magnetic axis,
(3) the angle $\betaobs$ between the rotation axis and the observer's line of sight,
(4) the angular position $\thetaj$ of the $j$-bundle footprint, and 
(5) the angular width $\Delta \thetaj$ of the $j$-bundle footprint.
In addition, the reference point of the rotational phase,  $\phaseref$, is a free parameter, since we fit the phase-resolved spectra.

We follow the method presented in \citet{hascoet2014}, and explore the whole parameter space by fitting the phase-averaged spectrum of the total emission (pulsed+unpulsed) and phase-resolved spectra of the pulsed emission. In order to get sufficient photon statistics, we used only three phase bins: `A' (0.05--0.35), `B' (0.35--0.70), and `C' (0.70--1.05), roughly covering the primary pulse peak, the minima and the sub-peak, respectively.
\AB{The bins are indicated}
in the last panel of Figure~\ref{fig:pulse_profile}. 
\AB{The phase bin with the lowest flux is assumed to represent the ``DC'' (unpulsed)
component; its spectrum is subtracted from the total spectrum in each phase bin to 
obtain the spectrum of the pulsed component. The}
\nustar\ data are fitted above 16\,keV, where the hard component becomes 
dominant and the coronal outflow model has to account for most of the X-ray emission. 

\begin{figure*}
\begin{center}
\begin{tabular}{cccc} 
\ \ \ \ \large {$p$-value} & \\
\vspace{-0.5cm}
\includegraphics[trim = 7.6cm 4.1cm 8.cm 0.8cm, clip, height=0.4\textwidth]{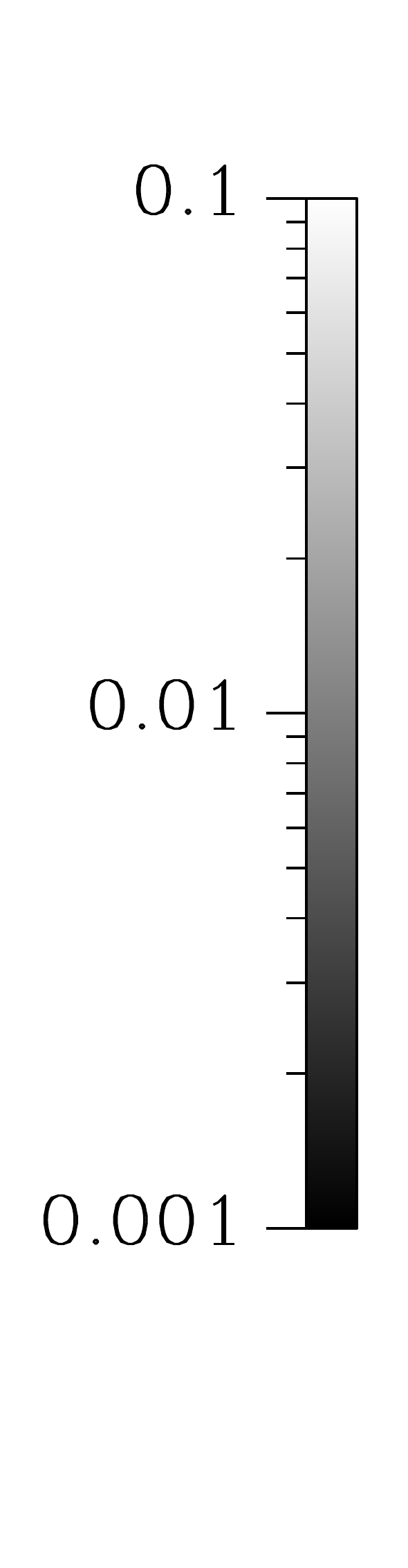}  
& \hspace{-0.2cm}\includegraphics[trim = 0.7cm 4.0cm 2.0cm 0.7cm, clip, height=0.4\textwidth]{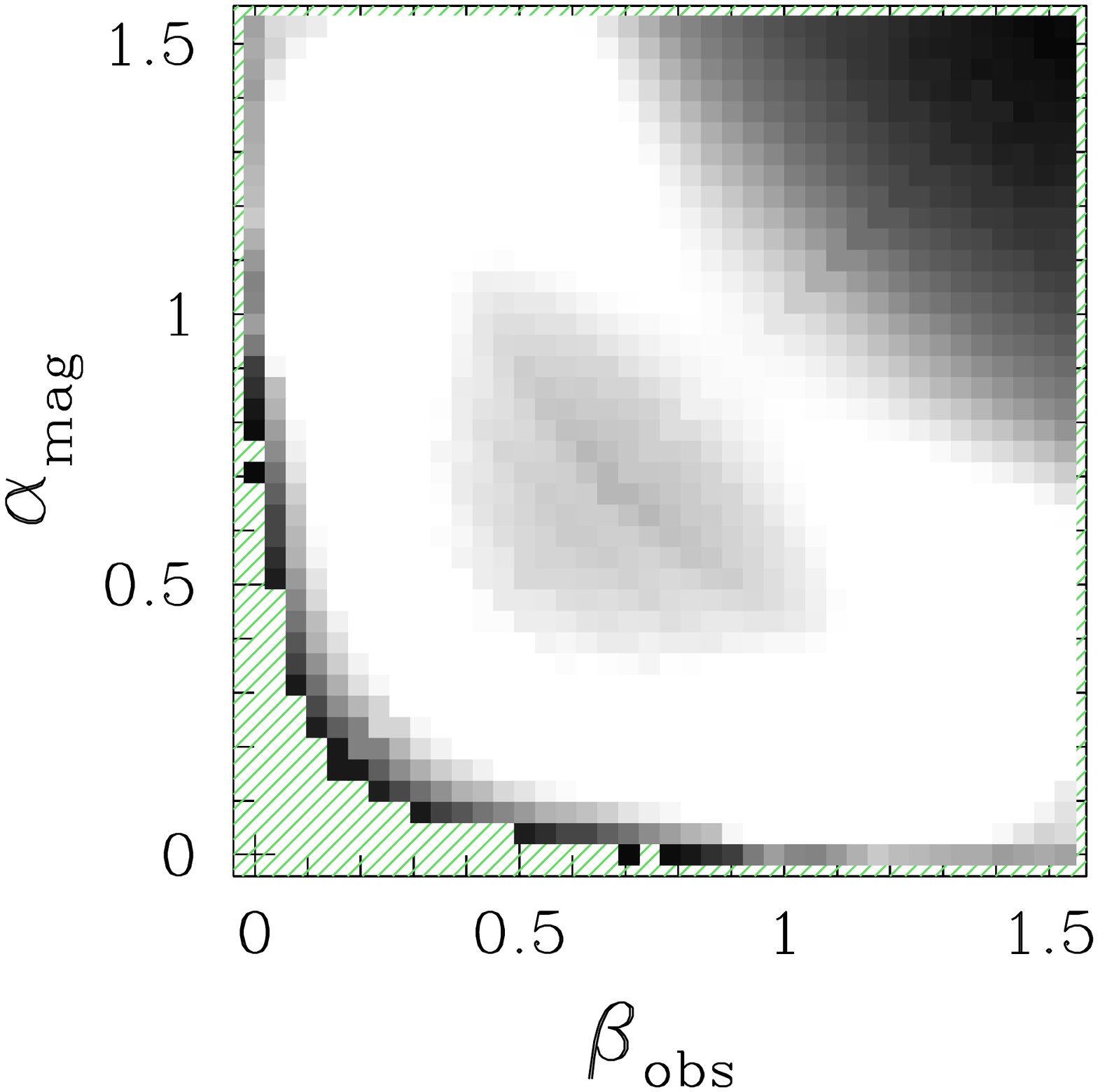}
& \hspace{-0.2cm}\includegraphics[trim = 4.6cm 4.0cm 2.0cm 0.7cm, clip, height=0.4\textwidth]{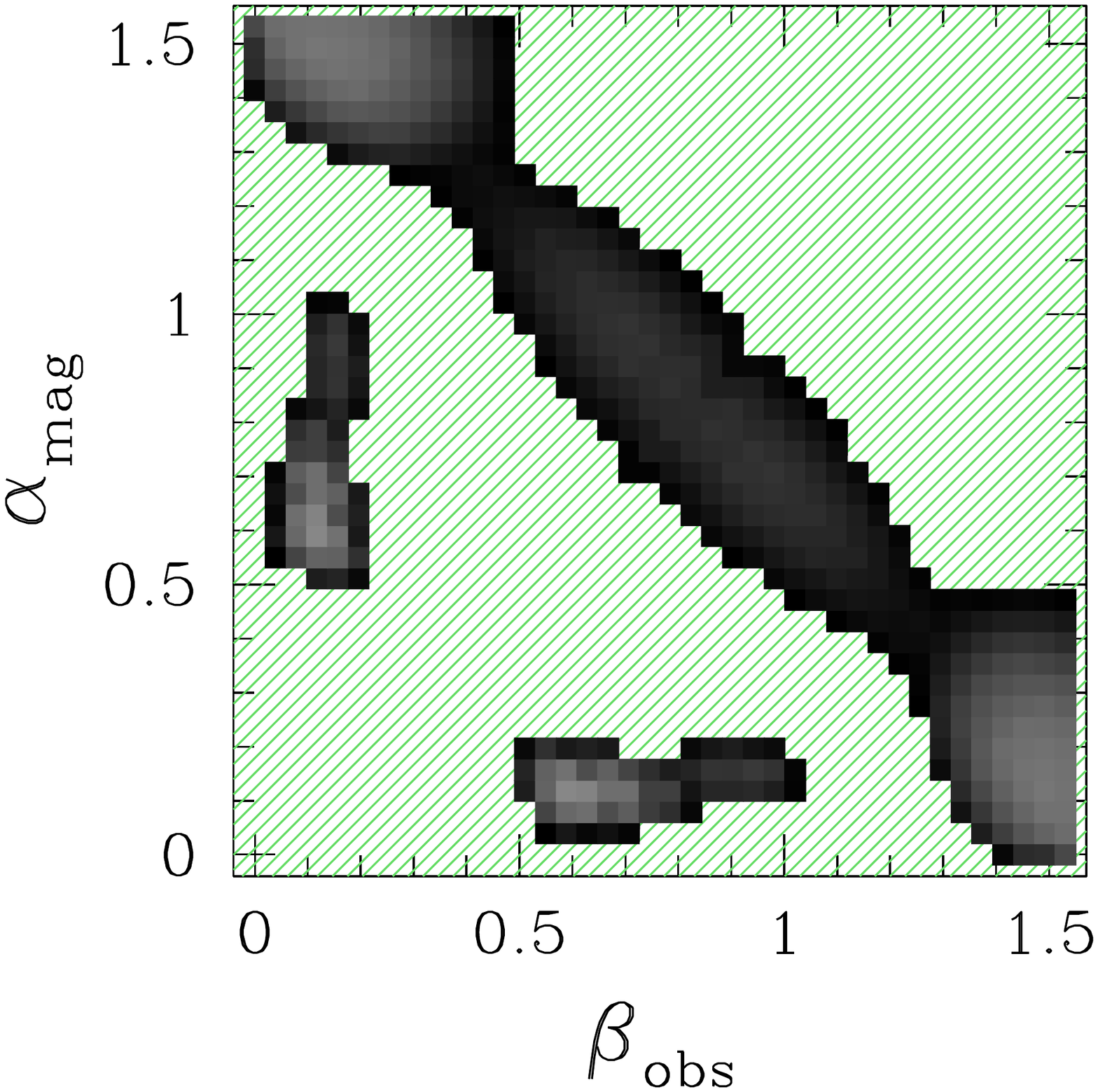}
& \hspace{-0.2cm}\includegraphics[trim = 4.6cm 4.0cm 2.0cm 0.7cm, clip, height=0.4\textwidth]{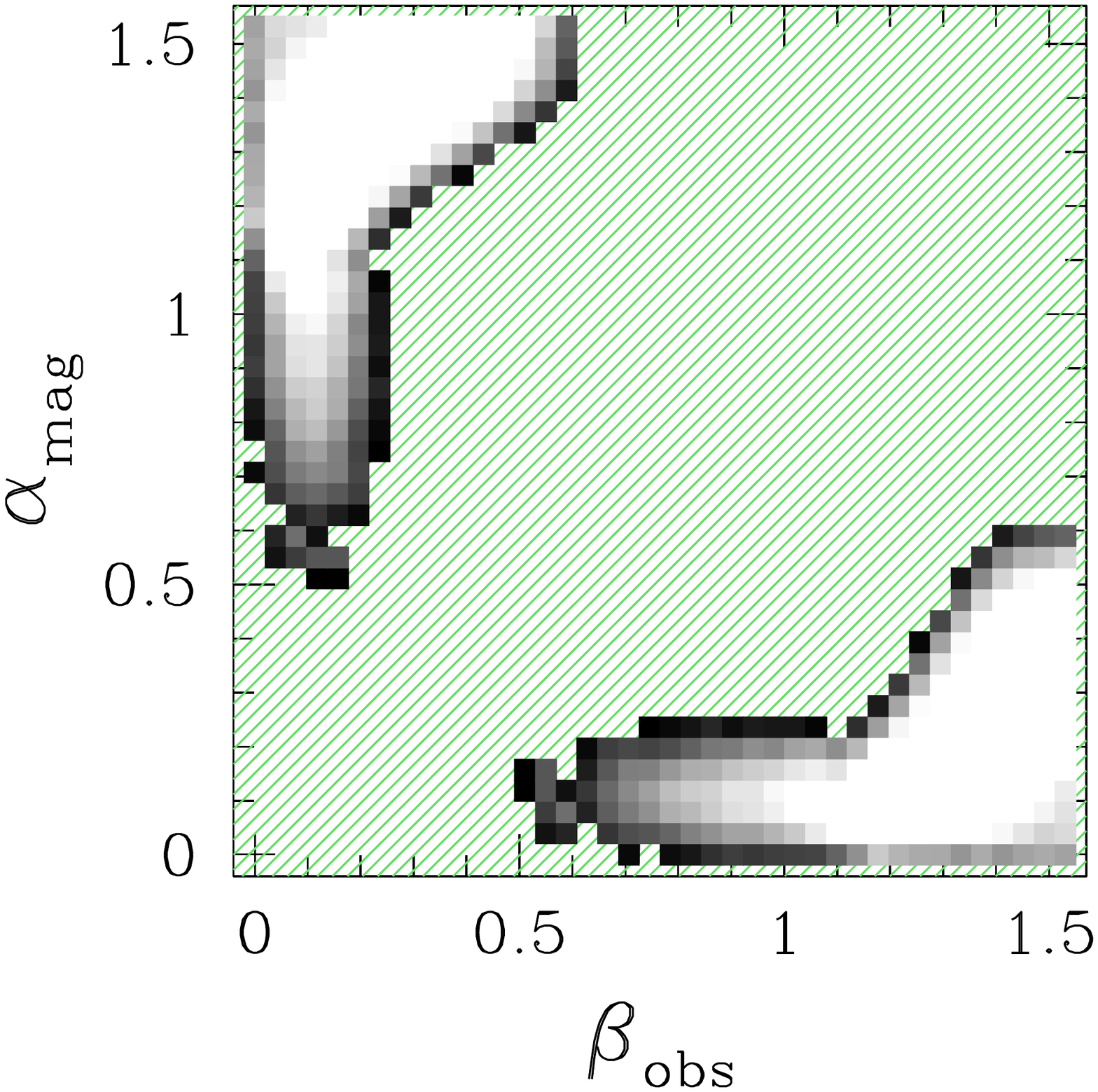} 
\end{tabular}
\end{center}
\caption{Maps of $p$-values for the fit of the hard X-ray component with the coronal outflow model; the $p$-values are shown in the plane of ($\alpha_{\rm mag}$, $\beta_{\rm obs}$) and maximized over the other parameters. The $\alpha_{\rm mag}$ axis is common for all the plots. The $p$-value scale is shown on the left. The hatched green regions have $p$-values smaller than $0.001$; the white regions have $p$-values greater than $0.1$. Interchanging the values of $\alphamag$ and $\betaobs$ does not change the model spectrum, as long as the $j$-bundle is assumed to be axisymmetric. Therefore, the map of $p$-values is symmetric about the line of $\betaobs=\alphamag$. \textit{Left:} $p$-value map when $\Delta \thetaj$ is thawed as a free parameter. \textit{Middle:} $p$-value map when the footprint width is frozen to $\Delta \thetaj = \thetaj / 2$. \textit{Right:} $p$-value map when the footprint area of the $j$-bundle, $\mathcal{A}_{j}$, is restricted to be in the interval $ 2.5\times10^{-3} < \mathcal{A}_{j} / \mathcal{A}_{\rm NS} < 10^{-2}$ (see discussion).}
\label{fig_pvalue}
\end{figure*}

The left panel of Figure \ref{fig_pvalue} shows the map of $p$-values in the plane $(\alphamag , \betaobs)$. The parameter space appears to be largely degenerate. For comparison with the results of \citet{hascoet2014} (discussed further in Section~\ref{sect_discussion} below), we also show the resulting $p$-value map when the footprint width is fixed to be $\Delta \thetaj  = \thetaj / 2$,
\AB{i.e. thin rings are excluded.}
\AB{Then the}
 degeneracy of the parameter space is 
\AB{significantly}
reduced,
\AB{and the results are consistent with those of Hasco\"et et al. (2014).}

Using the obtained best-fit model for the hard X-ray component, we have investigated the remaining soft X-ray component. The procedure is similar to that in \citet{hascoet2014}: we freeze the best-fit parameters of the outflow model, and fit the spectrum in the 0.5--79\,keV band \revision{including} the \swift-XRT data. As in \citet{hascoet2014}, we find that the 
spectrum is well fitted by the sum of one blackbody, one modified blackbody\footnote{In 
     this model, dubbed $BB_{\rm tail}$ in \citet{vogel2014}, the Wien tail of the blackbody is   
     replaced by a power-law ``smoothly'' connected at the photon energy $E_{\rm tail}$.   
     Here ``smoothly'' means 
     \AB{that}
     the photon spectrum and its derivative are continuous at  $E_{\rm tail}$.} 
and the coronal outflow emission (which dominates above 10\,keV). The (cold) blackbody and the (hot) modified blackbody have luminosities $L_{c} = 2.5(3)\times10^{35}\,\mathrm{erg\,s^{-1}}$, $L_{h} = 3.33(4)\times10^{34}\,\mathrm{erg\,s^{-1}}$ and temperatures $kT_{c} = 0.408(3)$\,keV, $kT_{h} = 0.85(1)$\,keV similar to those fit by model II in Section~\ref{sec:spectra_phase_average}. The power-law tail of the modified hot blackbody starts at $E_{\rm tail} = 5.7(1)$\,keV.

\section{Discussion and Conclusion}
\label{sec:discussion}
We have described timing and spectral analysis of simultaneous 0.3--79\,keV \swift-XRT and \nustar\ observations of \axp. Using Fourier analysis we present the variation in pulse shape and pulse fraction over the soft X-ray and hard X-ray bands. 
We find a significant change in pulse structure at the cross-over between the soft-energy peak where the modified blackbody emission is dominant and the hard-energy peak, where the magnetospheric tail emission is dominant. We do not find evidence for phase modulation in the 15--40\,keV lightcurve as reported by \citet{makishima2014}. We find that the phase-averaged spectrum is best modeled by a phenomenological \texttt{nthcomp}+PL model. The BB+2PL and 2BB+PL models that were traditionally used to fit the data do not provide statistically acceptable descriptions. 
Fitting the phase-resolved \swift-XRT and \nustar\ spectra of \axp, we find that the spectral shape parameters do not show statistically significant variations compared to the phase-averaged fits. However, the normalizations of the spectral components vary significantly at phases corresponding to peaks and dips in the pulse profiles. Finally, we place constraints on the geometry of \axp\ using the electron-position outflow models of \citet{beloborodov2013b}.

\subsection{Timing Analysis}
The low-energy pulse shapes measured from \swift-XRT and \nustar\ agree with the measurements of dH08 gathered with \xmm. In particular, the pulse profiles bear remarkable similarity with the data gathered on 2004 July 25 (dataset C) and on 2004 March 1 (dataset B) and are less similar to the previous observations (dataset A, gathered on 2003 January 04). The separation between the peaks in \swift-XRT ($\Delta\phi=0.35$, in 0.3--1.5\,keV) matches that in the \xmm\ data ($\Delta\phi=0.35$, in 0.8--2.0\,keV). The separation between the dips ($\Delta\phi=0.6$) and the relative pulse heights also match well between the two data sets. Similarly, the pulse shapes and relative heights between \nustar\ 3--5\,keV and 5--8\,keV profiles and the \xmm\ 2--8\,keV profiles are morphologically similar. 

Similarly, the \nustar\ 3--5\,keV profile agrees with the 2--4\,keV \emph{RXTE} pulse profiles of \citet{dib2007} obtained between 2005 March and 2006 February. However, there are increasing differences between the \nustar\ profile and the \emph{RXTE} pulse profiles at epochs going backward from 2005 to 1996. The 6--8\,keV profile between March 2005 and February 2006 shows a slightly broader main peak than the 5--8\,keV \nustar\ pulse profiles. Our low-energy pulse shapes agree well with 0.5--2\,keV and 2--10\,keV \xmm\ pulse profiles of \citet{gonzalez2010} obtained between 2006 July and 2008 March with the match improving as the compared epochs become closer.

We find slight differences between the \nustar\ 20--35\,keV and 35--50\,keV profiles and \textit{INTEGRAL} 20--50\,keV profiles described in dH08. The \nustar\ profiles show a primary peak (at $\phi=0.2$) that is 40\% higher than the secondary peak (at $\phi=0.7$). In the \textit{INTEGRAL} profiles, the peak separations are similar ($\Delta\phi=0.53$) but the peak count rates were equal. Similarly, we find that the \nustar\ 50--79\,keV profiles show evidence of a double-peaked structure, with two sharp peaks separated by $\Delta\phi=0.3$. The corresponding 50--160\,keV \textit{INTEGRAL} profile shows a single-peaked structure. While it is possible that the pulse profile has changed, note that the 50--79\,keV band would contribute only 35\% of the photon flux as compared to the 50--160\,keV energy band for a power-law spectrum with $\Gamma=0.65$. Hence the difference in pulse profile may also be attributable to the difference in energy ranges. We also observe that the relative height of the primary pulse (at $\phi\approx0.3$) compared to the pulse at $\phi\approx0.8$ is decreasing with increasing energy through the 20--35\,keV, 35--50\,keV and 50--79\, keV plots. Hence it is not inconceivable that at energies higher than 79\,keV, the pulse at $\phi\approx0.8$ starts to dominate the pulse profile.

\subsubsection{Non-Detection of Precession}
After repeating the analysis steps of \citet{makishima2014} on 15--40\,keV \nustar\ data of \axp, we did not detect any phase modulation that can be interpreted as precession of the neutron star. The pulse profile from the \nustar\ data, while very consistent in shape and amplitude with the double-peaked profiles of dH08, are very different in shape and amplitude from the triple-peaked profiles of ME14 obtained after phase-demodulation. It is possible that the precession signal may be time-varying, having been detected in 2009 but not in 2007 and 2014. However, considering the necessary reconfiguration in the neutron star moments of inertia ($\Delta I/I\sim10^{-4}$) and the corresponding reconfiguration of a $10^{16}$\,G toroidal magnetic field, it is surprising that the timing ephemeris, rotational spin-down and pulse profiles remain consistent between 2007 and 2014. Further searches of phase modulation will help to confirm and understand the mechanics of this result. 

\subsubsection{Comparisons with Other Magnetars}
The trend of pulse fraction as a function of energy varies from magnetar to magnetar though many have a pulse fraction increasing with energy \citep[see for example][]{kuiper2006, denhartog2008a, denhartog2008b}. In \axp\ we observe that $PF_\mathrm{RMS}$ increases up to a value of 20\% and possibly shows a small decline towards 40\,keV or possibly stays constant at $\approx$20\%. In 1E\,2259+596, $PF_\mathrm{RMS}$ was seen to monotonically rise to approximately 70\% at 20\,keV \citep{vogel2014} and in 1E\,1841$-$045, $PF_\mathrm{RMS}$ was seen to rise to a value of approximately 17\% at 10\,keV, decrease to 12\% at 20\,keV and rise again to approximately 17--20\% between 30--79\,keV \citep{an2013,an2015}. At the same time, $PF_\mathrm{area}$ was measured to increase from 25\% at 1--2\,keV and increase to 50\% at 50\,keV. For 1RXS\,J170849$-$400910 \citep{denhartog2008b}, the pulse fraction (reported as $PF_\mathrm{area}$, not $PF_\mathrm{RMS}$) was shown to be nearly constant at approximately 40\% between an energy range from 0.7--200\,keV. In fact, the pulse fraction decreases slightly from about 50\% at 1\,keV to about 30\% at 3\,keV, rising back to about 40\% at higher energies.

The X-ray pulse profiles of magnetars, affected by the geometry of the magnetic field and rotation axis, are similarly diverse. The pulse profiles of \axp\ are primarily double peaked for most energy bands with each pulse width being $\delta\phi\approx0.25$. Compared to these, the pulse profile of 1E\,1841$-$045 \citep{an2013,an2015} is comprised of large, single-peaked humps that are about $\delta\phi\approx0.75$ wide (except for the double-peaked structure emerging between 23.8--35.2\,keV). The pulse profiles of 1RXS\,J170849$-$400910 \citep{denhartog2008b} are dominated by a single pulse peak with a width $\delta\phi=0.35$ at most energy bands; however, there is a distinct shift between pulse positions below and above 8\,keV, suggesting that the dominant emission mechanism changes drastically. The pulse profiles of 1E\,2259+586 show complicated structure, with narrow peaks ($\delta\phi\approx0.25$) that can possibly shift slightly with energy \citep{vogel2014}. The pulse profiles of \axp\ are therefore morphologically more similar to those of 1RXS\,J170849$-$400910 than those from 1E\,2259+586 and 1E\,1841$-$045. 

A possible source for these differences may be the size and geometry of the hot-spot emitting area on each magnetar. Comparing the size of the $j$-bundle, $\theta_j$ in the outflow model fits (see Section~\ref{sect_discussion}) suggests a rough pattern, albeit in a very limited sample size. For 1E\,1841$-$045, the magnetar with the broadest pulse profiles, $\theta_j\lesssim 0.4$\,rad \citep{an2013,an2015} while for 1RXS\,J170849$-$400910 and \axp\ with narrower pulse profiles, $\theta_j<0.15$\,rad and $<0.23$\,rad, respectively \citep{hascoet2014}. The outflow model fit for 1E\,2259+586, which shows a complicated narrow pulse profile, statistically prefers a complicated ring-shaped $j$-bundle with $0.4\,\mathrm{rad}<\theta_j<0.75\,\mathrm{rad}$ and $\Delta\theta_j/\theta_j$, the ring-width fracion, $<0.2$ \citep{vogel2014}.     

\subsection{Spectral Analysis}
We fit different spectral models to the soft and hard energy spectra from \swift-XRT and \nustar. We find that the 2BB+PL and BB+2PL models do not fit the cross-over region (approximately 5--15\,keV) of the spectrum well. This causes a distortion in the fitting of the hard power-law and a residual is left at the high energies ($>50$\,keV). The \texttt{nthcomp}+PL model provides a statistically better fit than the two models, especially for the cross-over region. The hard PL index $\Gamma_H$ measured from this model best matches the $\Gamma_H=0.65\pm0.09$ measured after restricting the energy range to be between 20--79\,keV. 

We find that the spectral turnover $\Gamma_S-\Gamma_H=3.56$ (BB+2PL model) and $\Gamma_S-\Gamma_H=4.11$ (\texttt{nthcomp}+PL) are higher than the values reported for the total flux $\Gamma_S-\Gamma_H=2.6$ reported by \citet{kaspi2010}. Using the independent value of $\Gamma_H=0.65\pm0.09$ slightly increases the discrepancy. Placing these values on the $\Gamma_S-\Gamma_H$ vs $\log(B/10^{14}\,\mathrm{G})$ plot \citep[as shown in][]{vogel2014}, does not change the observed decreasing trend between  $\Gamma_S-\Gamma_H$ and $\log(B/10^{14}\,\mathrm{G})$. 

Fitting the same models to the phase-resolved spectra shows that the spectral shape parameters ($\Gamma_H$, $k_BT_\mathrm{BB}$ and $\Gamma_S$) are consistent within 3-$\sigma$ error bars to the values measured from phase-averaged spectral fits. However, the normalization of the hard X-ray and soft X-ray components varies significantly as a function of phase. From Figure~\ref{fig:spectra_phase_resolved}, we can identify the increase in the hard power-law normalization in the 0.2--0.4 phase range with the peak in the 20--35\,keV and 35--50\,keV pulse profiles and the dip in the soft X-ray components normalization (\texttt{nthcomp} or soft power law, depending on the model fits) with the dip at phase $\phi=0.5$ in the 3--5\,keV and 5--8\,keV bands. This suggests a clear differentiation between the low-energy and high-energy spectral components.

We find that the best-fitting \texttt{nthcomp}+PL model yields $N_\mathrm{H}=(6.5\pm0.2)\times10^{21}\,\mathrm{cm^{-2}}$, consistent with that measured by D06b and also consistent with later broadband fits by \citet{enoto2011}.

\subsection{Outflow Model}
\label{sect_discussion}
We find that the coronal outflow model provides consistent fits to the phase-resolved \nustar\ spectra of \fouru. \citet{hascoet2014} obtained a similar conclusion by fitting the {\em INTEGRAL} phase-resolved spectra  of \citet{denhartog2008a}.
\AB{Their model assumed that the outflow occurs along magnetic field lines emerging 
from a polar cap on the star or a thick ring $\Delta \thetaj = \thetaj / 2$. 
An excellent fit was provided by this model in a small region of parameter space, giving strong constraints on $\alpha_{\rm mag}$ 
and $\beta_{\rm obs}$.}
\AB{Motivated by the recent analysis of 1E\,2259$+$586  \citep{vogel2014}, 
we explored a more general outflow model that allows the footprint of $j$-bundle to be 
a ring of arbitrary thickness $\Delta \thetaj$. 
We found that a thin-ring configuration is also able to fit the phase-resolved
spectrum of \axp, and in a broader range of parameters. This degeneracy is absent 
in 1E\,1841$-$045, where only a thick ring or a polar cap is allowed \citep{an2015}.} 

The soft X-ray component (below $\sim 10$\,keV) is well fitted by the sum of one cold blackbody and one modified hot blackbody. The cold blackbody covers a large fraction of the neutron star area,  $\mathcal{A}_{c} \approx 0.7\,\mathcal{A}_{\rm NS}$. The emission area of the hot blackbody is small, $\mathcal{A}_{h} \approx 0.005\,\mathcal{A}_{\rm NS}$. In the coronal outflow model, the footprint of the $j$-bundle is expected to form a hot spot, as some particles accelerated in the $j$-bundle flow back to the neutron star and bombard its surface. If the modified hot blackbody is interpreted as the thermal emission from the footprint, then the measured $\mathcal{A}_{h}$ can be used  as a constraint on the footprint of the active $j$-bundle. The right panel of Figure~\ref{fig_pvalue} shows the {\em NuSTAR} $p$-value when the $j$-bundle footprint area, $\mathcal{A}_{j} = \pi \sin^2 \thetaj$, is restricted to be between $\mathcal{A}_{h} / 2 = 2.5 \times 10^{-3}\mathcal{A}_{\rm NS}$ and $\mathcal{A}_{h} \times 2 = 10^{-2}\mathcal{A}_{\rm NS}$. Then the degeneracy of the model is reduced and a broad region of the parameter space around the line $\alphamag = \betaobs$ becomes excluded. 

The outflow model predicts the X-ray flux below $\sim$1\,MeV to be dominated by photons polarized perpendicular to the magnetic field while an excess of parallel-polarized photons is expected through photon splitting at higher energies \citep{beloborodov2013b}. The model of magnetospheric emission will be crucially tested by future X-ray polarimetry instruments such as \emph{ASTRO-H}-SGD \citep[][ during flares]{tajima2010}, \emph{ASTROSAT}-CZTI \citep{chattopadhyay2014}, \emph{POLAR} \citep{produit2005} and \emph{X-Calibur} \citep{beilicke2014}.

In conclusion, we have presented a timing and spectral analysis of simultaneous 0.5--79\,keV observations of \axp\ using \swift-XRT and \nustar. The rotational period of \axp\ is consistent with that expected from extrapolation of the timing solution since the last glitch \citep{dib2014}. We have not detected the 55-ks time period, 0.7-s amplitude phase modulation in the 15--40\,keV \textit{Suzaku}-HXD data from 2007 reported by \citet{makishima2014} that was ascribed to the free-precession of \axp. While this precession may be time-varying, the consistency of the rotational ephemeris and pulse profile between 2007 and 2014 needs to be explained. We have shown that the pulse profile changes character (dominance of the first harmonic vs the second harmonic) at around 30\,keV. While the low-energy pulse profiles were consistent with previously presented pulse profiles \citep[between 2006 and 2008:][]{dib2007,gonzalez2010}, we have observed morphological differences between the hard energy pulse profiles of \nustar\ and \textit{INTEGRAL}. We have shown that the RMS pulse fraction has an increasing trend with energy, reaching a value of up to 20\%, however, it shows some evidence of a decrease at about 40\,keV similar to that observed in 1E\,1841$-$045 and contrary to the smooth increase of pulse fraction in 1E\,2259+586 that increases to nearly 80\%.

We have shown that the energy spectrum of \axp\ between 0.5--79\,keV is better described by a Comptonized blackbody + hard PL model than the previously used BB+2PL or 2BB+PL models with a hard power-law ($\Gamma_H=0.65\pm0.09$) dominating the spectrum above 20\,keV. The low-energy spectrum ($<10$\,keV) may still be fit with the BB+PL model, however, this model cannot fit the observed spectrum between 10--20\,keV.

We have fitted the phase-resolved spectra of \axp\ with the $e^\pm$ outflow model of \citet{beloborodov2013a}
using the analysis method of \citet{hascoet2014}. Our results show that the outflow model gives a consistent physical description of the phase-resolved spectra, and the results are consistent with those derived from \textit{INTEGRAL} data. 
\AB{We found that significant degeneracy appears in the inferred parameters of the 
inclined rotator}
$\alpha_{\rm mag}$ and $\beta_{\rm obs}$
\AB{if the footprint of the $j$-bundle is allowed to be a thin ring.} 
The degeneracy is 
\AB{significantly}
reduced if the 
\AB{footprint area}
  $A_j$ is restricted to be similar to the area of the blackbody hotspot that covers 0.5\% of the neutron star surface.

\acknowledgements
This work was supported under NASA Contract No. NNG08FD60C, and made use of data from the \nustar\ mission, a project led by the California Institute of Technology, managed by the Jet Propulsion Laboratory, and funded by the National Aeronautics and Space Administration. We thank the \nustar\ Operations, Software and Calibration teams for support with the execution and analysis of these observations. This research has made use of the \nustar\ Data Analysis Software (\textit{NuSTARDAS}) jointly developed by the ASI Science Data Center (ASDC, Italy) and the California Institute of Technology (USA). V.M.K. acknowledges support from an NSERC Discovery Grant and Accelerator Supplement, the FQRNT Centre de Recherche Astrophysique du Qu\'{e}bec, an R. Howard Webster Foundation Fellowship from the Canadian Institute for Advanced Research (CIFAR), the Canada Research Chairs Program and the Lorne Trottier Chair in Astrophysics and Cosmology. 

\bibliographystyle{apj}
\bibliography{paper}

\begin{thebibliography}{67}
\expandafter\ifx\csname natexlab\endcsname\relax\def\natexlab#1{#1}\fi

\bibitem[{{An} {et~al.}(2013){An}, {Hasco{\"e}t}, {Kaspi}, {Beloborodov},
  {Dufour}, {Gotthelf}, {Archibald}, {Bachetti}, {Boggs}, {Christensen},
  {Craig}, {Greffenstette}, {Hailey}, {Harrison}, {Kitaguchi}, {Kouveliotou},
  {Madsen}, {Markwardt}, {Stern}, {Vogel}, \& {Zhang}}]{an2013}
{An}, H., {Hasco{\"e}t}, R., {Kaspi}, V.~M., {et~al.} 2013, \apj, 779, 163

\bibitem[{{An} {et~al.}(2014{\natexlab{a}}){An}, {Kaspi}, {Beloborodov},
  {Kouveliotou}, {Archibald}, {Boggs}, {Christensen}, {Craig}, {Gotthelf},
  {Grefenstette}, {Hailey}, {Harrison}, {Madsen}, {Mori}, {Stern}, \&
  {Zhang}}]{an2014b}
{An}, H., {Kaspi}, V.~M., {Beloborodov}, A.~M., {et~al.} 2014{\natexlab{a}},
  \apj, 790, 60

\bibitem[{{An} {et~al.}(2014{\natexlab{b}}){An}, {Kaspi}, {Archibald},
  {Bachetti}, {Bhalerao}, {Bellm}, {Beloborodov}, {Boggs}, {Chakrabarty},
  {Christensen}, {Craig}, {Dufour}, {Forster}, {Gotthelf}, {Grefenstette},
  {Hailey}, {Harrison}, {Hasco{\"e}t}, {Kitaguchi}, {Kouveliotou}, {Madsen},
  {Mori}, {Pivovaroff}, {Rana}, {Stern}, {Tendulkar}, {Tomsick}, {Vogel},
  {Zhang}, \& {NuSTAR Team}}]{an2014}
{An}, H., {Kaspi}, V.~M., {Archibald}, R., {et~al.} 2014{\natexlab{b}},
  Astronomische Nachrichten, 335, 280

\bibitem[{{An} {et~al.}(2015){An}, {Archibald}, {Hascoet}, {Kaspi},
  {Beloborodov}, {Archibald}, {Beardmore}, {Boggs}, {Christensen}, {Craig},
  {Gehrels}, {Hailey}, {Harrison}, {Kennea}, {Kouveliotou}, {Stern}, {Younes},
  \& {Zhang}}]{an2015}
{An}, H., {Archibald}, R.~F., {Hascoet}, R., {et~al.} 2015, ArXiv e-prints,
  1505.03570

\bibitem[{{Anders} \& {Ebihara}(1982)}]{anders1982}
{Anders}, E., \& {Ebihara}, M. 1982, \gca, 46, 2363

\bibitem[{{Anders} \& {Grevesse}(1989)}]{anders1989}
{Anders}, E., \& {Grevesse}, N. 1989, \gca, 53, 197

\bibitem[{{Arnaud}(1996)}]{arnaud1996}
{Arnaud}, K.~A. 1996, in Astronomical Society of the Pacific Conference Series,
  Vol. 101, Astronomical Data Analysis Software and Systems V, ed. G.~H.
  {Jacoby} \& J.~{Barnes}, 17

\bibitem[{{Asplund} {et~al.}(2005){Asplund}, {Grevesse}, \&
  {Sauval}}]{asplund2005}
{Asplund}, M., {Grevesse}, N., \& {Sauval}, A.~J. 2005, in Astronomical Society
  of the Pacific Conference Series, Vol. 336, Cosmic Abundances as Records of
  Stellar Evolution and Nucleosynthesis, ed. T.~G. {Barnes}, III \& F.~N.
  {Bash}, 25

\bibitem[{{Asplund} {et~al.}(2009){Asplund}, {Grevesse}, {Sauval}, \&
  {Scott}}]{asplund2009}
{Asplund}, M., {Grevesse}, N., {Sauval}, A.~J., \& {Scott}, P. 2009, \araa, 47,
  481

\bibitem[{{Balucinska-Church} \& {McCammon}(1992)}]{balucinskachurch1992}
{Balucinska-Church}, M., \& {McCammon}, D. 1992, \apj, 400, 699

\bibitem[{{Beilicke} {et~al.}(2014){Beilicke}, {Kislat}, {Zajczyk}, {Guo},
  {Endsley}, {Stork}, {Cowsik}, {Dowkontt}, {Barthelmy}, {Hams}, {Okajima},
  {Sasaki}, {Zeiger}, {de Geronimo}, {Baring}, \& {Krawczynski}}]{beilicke2014}
{Beilicke}, M., {Kislat}, F., {Zajczyk}, A., {et~al.} 2014, Journal of
  Astronomical Instrumentation, 3, 40008

\bibitem[{{Beloborodov}(2009)}]{beloborodov2009}
{Beloborodov}, A.~M. 2009, \apj, 703, 1044

\bibitem[{{Beloborodov}(2013{\natexlab{a}})}]{beloborodov2013b}
---. 2013{\natexlab{a}}, \apj, 777, 114

\bibitem[{{Beloborodov}(2013{\natexlab{b}})}]{beloborodov2013a}
---. 2013{\natexlab{b}}, \apj, 762, 13

\bibitem[{{Brazier}(1994)}]{brazier1994}
{Brazier}, K.~T.~S. 1994, \mnras, 268, 709

\bibitem[{{Burrows} {et~al.}(2005){Burrows}, {Hill}, {Nousek}, {Kennea},
  {Wells}, {Osborne}, {Abbey}, {Beardmore}, {Mukerjee}, {Short}, {Chincarini},
  {Campana}, {Citterio}, {Moretti}, {Pagani}, {Tagliaferri}, {Giommi},
  {Capalbi}, {Tamburelli}, {Angelini}, {Cusumano}, {Br{\"a}uninger}, {Burkert},
  \& {Hartner}}]{burrows2005}
{Burrows}, D.~N., {Hill}, J.~E., {Nousek}, J.~A., {et~al.} 2005, \ssr, 120, 165

\bibitem[{{Chattopadhyay} {et~al.}(2014){Chattopadhyay}, {Vadawale}, {Rao},
  {Sreekumar}, \& {Bhattacharya}}]{chattopadhyay2014}
{Chattopadhyay}, T., {Vadawale}, S.~V., {Rao}, A.~R., {Sreekumar}, S., \&
  {Bhattacharya}, D. 2014, Experimental Astronomy, 37, 555

\bibitem[{{den Hartog} {et~al.}(2006){den Hartog}, {Hermsen}, {Kuiper}, {Vink},
  {in't Zand}, \& {Collmar}}]{denhartog2006}
{den Hartog}, P.~R., {Hermsen}, W., {Kuiper}, L., {et~al.} 2006, \aap, 451, 587

\bibitem[{{den Hartog} {et~al.}(2008{\natexlab{a}}){den Hartog}, {Kuiper}, \&
  {Hermsen}}]{denhartog2008b}
{den Hartog}, P.~R., {Kuiper}, L., \& {Hermsen}, W. 2008{\natexlab{a}}, \aap,
  489, 263

\bibitem[{{den Hartog} {et~al.}(2008{\natexlab{b}}){den Hartog}, {Kuiper},
  {Hermsen}, {Kaspi}, {Dib}, {Kn{\"o}dlseder}, \& {Gavriil}}]{denhartog2008a}
{den Hartog}, P.~R., {Kuiper}, L., {Hermsen}, W., {et~al.} 2008{\natexlab{b}},
  \aap, 489, 245

\bibitem[{{den Hartog} {et~al.}(2004){den Hartog}, {Kuiper}, {Hermsen}, \&
  {Vink}}]{denhartog2004ATel}
{den Hartog}, P.~R., {Kuiper}, L., {Hermsen}, W., \& {Vink}, J. 2004, The
  Astronomer's Telegram, 293, 1

\bibitem[{{Dib} \& {Kaspi}(2014)}]{dib2014}
{Dib}, R., \& {Kaspi}, V.~M. 2014, \apj, 784, 37

\bibitem[{{Dib} {et~al.}(2007){Dib}, {Kaspi}, \& {Gavriil}}]{dib2007}
{Dib}, R., {Kaspi}, V.~M., \& {Gavriil}, F.~P. 2007, \apj, 666, 1152

\bibitem[{{Durant} \& {van Kerkwijk}(2006{\natexlab{a}})}]{durant2006a}
{Durant}, M., \& {van Kerkwijk}, M.~H. 2006{\natexlab{a}}, \apj, 650, 1070

\bibitem[{{Durant} \& {van Kerkwijk}(2006{\natexlab{b}})}]{durant2006b}
---. 2006{\natexlab{b}}, \apj, 650, 1082

\bibitem[{{Enoto} {et~al.}(2011){Enoto}, {Makishima}, {Nakazawa}, {Kokubun},
  {Kawaharada}, {Kotoku}, \& {Shibazaki}}]{enoto2011}
{Enoto}, T., {Makishima}, K., {Nakazawa}, K., {et~al.} 2011, \pasj, 63, 387

\bibitem[{{Enoto} {et~al.}(2010){Enoto}, {Nakazawa}, {Makishima}, {Rea},
  {Hurley}, \& {Shibata}}]{enoto2010}
{Enoto}, T., {Nakazawa}, K., {Makishima}, K., {et~al.} 2010, \apjl, 722, L162

\bibitem[{{Feldman}(1992)}]{feldman1992}
{Feldman}, U. 1992, \physscr, 46, 202

\bibitem[{{Giacconi} {et~al.}(1972){Giacconi}, {Murray}, {Gursky}, {Kellogg},
  {Schreier}, \& {Tananbaum}}]{giacconi1972}
{Giacconi}, R., {Murray}, S., {Gursky}, H., {et~al.} 1972, \apj, 178, 281

\bibitem[{{G{\"o}hler} {et~al.}(2004){G{\"o}hler}, {Staubert}, \&
  {Wilms}}]{gohler2004}
{G{\"o}hler}, E., {Staubert}, R., \& {Wilms}, J. 2004, \memsai, 75, 464

\bibitem[{{G{\"o}hler} {et~al.}(2005){G{\"o}hler}, {Wilms}, \&
  {Staubert}}]{gohler2005}
{G{\"o}hler}, E., {Wilms}, J., \& {Staubert}, R. 2005, \aap, 433, 1079

\bibitem[{{Gonzalez} {et~al.}(2010){Gonzalez}, {Dib}, {Kaspi}, {Woods}, {Tam},
  \& {Gavriil}}]{gonzalez2010}
{Gonzalez}, M.~E., {Dib}, R., {Kaspi}, V.~M., {et~al.} 2010, \apj, 716, 1345

\bibitem[{{Grevesse} \& {Sauval}(1998)}]{grevesse1998}
{Grevesse}, N., \& {Sauval}, A.~J. 1998, \ssr, 85, 161

\bibitem[{{Harrison} {et~al.}(2013){Harrison}, {Craig}, {Christensen},
  {Hailey}, {Zhang}, {Boggs}, {Stern}, {Cook}, {Forster}, {Giommi},
  {Grefenstette}, {Kim}, {Kitaguchi}, {Koglin}, {Madsen}, {Mao}, {Miyasaka},
  {Mori}, {Perri}, {Pivovaroff}, {Puccetti}, {Rana}, {Westergaard}, {Willis},
  {Zoglauer}, {An}, {Bachetti}, {Barri{\`e}re}, {Bellm}, {Bhalerao},
  {Brejnholt}, {Fuerst}, {Liebe}, {Markwardt}, {Nynka}, {Vogel}, {Walton},
  {Wik}, {Alexander}, {Cominsky}, {Hornschemeier}, {Hornstrup}, {Kaspi},
  {Madejski}, {Matt}, {Molendi}, {Smith}, {Tomsick}, {Ajello}, {Ballantyne},
  {Balokovi{\'c}}, {Barret}, {Bauer}, {Blandford}, {Brandt}, {Brenneman},
  {Chiang}, {Chakrabarty}, {Chenevez}, {Comastri}, {Dufour}, {Elvis}, {Fabian},
  {Farrah}, {Fryer}, {Gotthelf}, {Grindlay}, {Helfand}, {Krivonos}, {Meier},
  {Miller}, {Natalucci}, {Ogle}, {Ofek}, {Ptak}, {Reynolds}, {Rigby},
  {Tagliaferri}, {Thorsett}, {Treister}, \& {Urry}}]{harrison2013}
{Harrison}, F.~A., {Craig}, W.~W., {Christensen}, F.~E., {et~al.} 2013, \apj,
  770, 103

\bibitem[{{Hasco{\"e}t} {et~al.}(2014){Hasco{\"e}t}, {Beloborodov}, \& {den
  Hartog}}]{hascoet2014}
{Hasco{\"e}t}, R., {Beloborodov}, A.~M., \& {den Hartog}, P.~R. 2014, \apjl,
  786, L1

\bibitem[{{Hulleman} {et~al.}(2004){Hulleman}, {van Kerkwijk}, \&
  {Kulkarni}}]{hulleman2004}
{Hulleman}, F., {van Kerkwijk}, M.~H., \& {Kulkarni}, S.~R. 2004, \aap, 416,
  1037

\bibitem[{{Israel} {et~al.}(1993){Israel}, {Mereghetti}, \&
  {Stella}}]{israel1993}
{Israel}, G.~L., {Mereghetti}, S., \& {Stella}, L. 1993, \iaucirc, 5889, 1

\bibitem[{{Israel} {et~al.}(1994){Israel}, {Mereghetti}, \&
  {Stella}}]{israel1994}
---. 1994, \apjl, 433, L25

\bibitem[{{Israel} {et~al.}(1999){Israel}, {Oosterbroek}, {Angelini},
  {Campana}, {Mereghetti}, {Parmar}, {Segreto}, {Stella}, {van Paradijs}, \&
  {White}}]{israel1999}
{Israel}, G.~L., {Oosterbroek}, T., {Angelini}, L., {et~al.} 1999, \aap, 346,
  929

\bibitem[{{Juett} {et~al.}(2002){Juett}, {Marshall}, {Chakrabarty}, \&
  {Schulz}}]{juett2002}
{Juett}, A.~M., {Marshall}, H.~L., {Chakrabarty}, D., \& {Schulz}, N.~S. 2002,
  \apjl, 568, L31

\bibitem[{{Kaspi} \& {Boydstun}(2010)}]{kaspi2010}
{Kaspi}, V.~M., \& {Boydstun}, K. 2010, \apjl, 710, L115

\bibitem[{{Kaspi} {et~al.}(2014){Kaspi}, {Archibald}, {Bhalerao}, {Dufour},
  {Gotthelf}, {An}, {Bachetti}, {Beloborodov}, {Boggs}, {Christensen}, {Craig},
  {Grefenstette}, {Hailey}, {Harrison}, {Kennea}, {Kouveliotou}, {Madsen},
  {Mori}, {Markwardt}, {Stern}, {Vogel}, \& {Zhang}}]{kaspi2014}
{Kaspi}, V.~M., {Archibald}, R.~F., {Bhalerao}, V., {et~al.} 2014, \apj, 786,
  84

\bibitem[{{Kuiper} {et~al.}(2006){Kuiper}, {Hermsen}, {den Hartog}, \&
  {Collmar}}]{kuiper2006}
{Kuiper}, L., {Hermsen}, W., {den Hartog}, P.~R., \& {Collmar}, W. 2006, \apj,
  645, 556

\bibitem[{{Leahy}(1987)}]{leahy1987}
{Leahy}, D.~A. 1987, \aap, 180, 275

\bibitem[{{Lodders}(2003)}]{lodders2003}
{Lodders}, K. 2003, \apj, 591, 1220

\bibitem[{{Madsen} {et~al.}(2015){Madsen}, {Harrison}, {Markwardt}, {An},
  {Grefenstette}, {Bachetti}, {Miyasaka}, {Kitaguchi}, {Bhalerao},
  {Christensen}, {Craig}, {Fuerst}, {Walton}, {Hailey}, {Rana}, {Stern},
  {Westergaard}, \& {Zhang}}]{madsen2015}
{Madsen}, K.~K., {Harrison}, F.~A., {Markwardt}, C., {et~al.} 2015, ArXiv
  e-prints, 1504.01672

\bibitem[{{Makishima} {et~al.}(2014){Makishima}, {Enoto}, {Hiraga}, {Nakano},
  {Nakazawa}, {Sakurai}, {Sasano}, \& {Murakami}}]{makishima2014}
{Makishima}, K., {Enoto}, T., {Hiraga}, J.~S., {et~al.} 2014, Physical Review
  Letters, 112, 171102

\bibitem[{{Mereghetti}(2008)}]{mereghetti2008}
{Mereghetti}, S. 2008, \aapr, 15, 225

\bibitem[{{Mori} {et~al.}(2013){Mori}, {Gotthelf}, {Zhang}, {An}, {Baganoff},
  {Barri{\`e}re}, {Beloborodov}, {Boggs}, {Christensen}, {Craig}, {Dufour},
  {Grefenstette}, {Hailey}, {Harrison}, {Hong}, {Kaspi}, {Kennea}, {Madsen},
  {Markwardt}, {Nynka}, {Stern}, {Tomsick}, \& {Zhang}}]{mori2013}
{Mori}, K., {Gotthelf}, E.~V., {Zhang}, S., {et~al.} 2013, \apjl, 770, L23

\bibitem[{{Olausen} \& {Kaspi}(2014)}]{olausen2014}
{Olausen}, S.~A., \& {Kaspi}, V.~M. 2014, \apjs, 212, 6

\bibitem[{{Patel} {et~al.}(2003){Patel}, {Kouveliotou}, {Woods}, {Tennant},
  {Weisskopf}, {Finger}, {Wilson}, {G{\"o}{\u g}{\"u}{\c s}}, {van der Klis},
  \& {Belloni}}]{patel2003}
{Patel}, S.~K., {Kouveliotou}, C., {Woods}, P.~M., {et~al.} 2003, \apj, 587,
  367

\bibitem[{{Paul} {et~al.}(2000){Paul}, {Kawasaki}, {Dotani}, \&
  {Nagase}}]{paul2000}
{Paul}, B., {Kawasaki}, M., {Dotani}, T., \& {Nagase}, F. 2000, \apj, 537, 319

\bibitem[{{Predehl} \& {Schmitt}(1995)}]{predehl1995}
{Predehl}, P., \& {Schmitt}, J.~H.~M.~M. 1995, \aap, 293, 889

\bibitem[{{Produit} {et~al.}(2005){Produit}, {Barao}, {Deluit}, {Hajdas},
  {Leluc}, {Pohl}, {Rapin}, {Vialle}, {Walter}, \& {Wigger}}]{produit2005}
{Produit}, N., {Barao}, F., {Deluit}, S., {et~al.} 2005, Nuclear Instruments
  and Methods in Physics Research A, 550, 616

\bibitem[{{Rea} \& {Esposito}(2011)}]{rea2011}
{Rea}, N., \& {Esposito}, P. 2011, in High-Energy Emission from Pulsars and
  their Systems, ed. D.~F. {Torres} \& N.~{Rea}, 247

\bibitem[{{Rea} {et~al.}(2007){Rea}, {Nichelli}, {Israel}, {Perna},
  {Oosterbroek}, {Parmar}, {Turolla}, {Campana}, {Stella}, {Zane}, \&
  {Angelini}}]{rea2007}
{Rea}, N., {Nichelli}, E., {Israel}, G.~L., {et~al.} 2007, \mnras, 381, 293

\bibitem[{{Rea} {et~al.}(2010){Rea}, {Esposito}, {Turolla}, {Israel}, {Zane},
  {Stella}, {Mereghetti}, {Tiengo}, {G{\"o}tz}, {G{\"o}{\u g}{\"u}{\c s}}, \&
  {Kouveliotou}}]{rea2010}
{Rea}, N., {Esposito}, P., {Turolla}, R., {et~al.} 2010, Science, 330, 944

\bibitem[{{Scholz} {et~al.}(2014){Scholz}, {Kaspi}, \& {Cumming}}]{scholz2014}
{Scholz}, P., {Kaspi}, V.~M., \& {Cumming}, A. 2014, \apj, 786, 62

\bibitem[{{Tajima} {et~al.}(2010){Tajima}, {Blandford}, {Enoto}, {Fukazawa},
  {Gilmore}, {Kamae}, {Kataoka}, {Kawaharada}, {Kokubun}, {Laurent}, {Lebrun},
  {Limousin}, {Madejski}, {Makishima}, {Mizuno}, {Nakazawa}, {Ohno}, {Ohta},
  {Sato}, {Sato}, {Takahashi}, {Takahashi}, {Tanaka}, {Tashiro}, {Terada},
  {Uchiyama}, {Watanabe}, {Yamaoka}, \& {Yonetoku}}]{tajima2010}
{Tajima}, H., {Blandford}, R., {Enoto}, T., {et~al.} 2010, in Society of
  Photo-Optical Instrumentation Engineers (SPIE) Conference Series, Vol. 7732,
  Society of Photo-Optical Instrumentation Engineers (SPIE) Conference Series,
  16

\bibitem[{{Thompson} \& {Duncan}(1995)}]{thompson1995}
{Thompson}, C., \& {Duncan}, R.~C. 1995, \mnras, 275, 255

\bibitem[{{Thompson} \& {Duncan}(1996)}]{thompson1996}
---. 1996, \apj, 473, 322

\bibitem[{{Thompson} {et~al.}(2002){Thompson}, {Lyutikov}, \&
  {Kulkarni}}]{thompson2002}
{Thompson}, C., {Lyutikov}, M., \& {Kulkarni}, S.~R. 2002, \apj, 574, 332

\bibitem[{{Vogel} {et~al.}(2014){Vogel}, {Hasco{\"e}t}, {Kaspi}, {An},
  {Archibald}, {Beloborodov}, {Boggs}, {Christensen}, {Craig}, {Gotthelf},
  {Grefenstette}, {Hailey}, {Harrison}, {Kennea}, {Madsen}, {Pivovaroff},
  {Stern}, \& {Zhang}}]{vogel2014}
{Vogel}, J.~K., {Hasco{\"e}t}, R., {Kaspi}, V.~M., {et~al.} 2014, \apj, 789, 75

\bibitem[{{White} {et~al.}(1996){White}, {Angelini}, {Ebisawa}, {Tanaka}, \&
  {Ghosh}}]{white1996}
{White}, N.~E., {Angelini}, L., {Ebisawa}, K., {Tanaka}, Y., \& {Ghosh}, P.
  1996, \apjl, 463, L83

\bibitem[{{Wilms} {et~al.}(2000){Wilms}, {Allen}, \& {McCray}}]{wilms2000}
{Wilms}, J., {Allen}, A., \& {McCray}, R. 2000, \apj, 542, 914

\bibitem[{{Yan} {et~al.}(1998){Yan}, {Sadeghpour}, \& {Dalgarno}}]{yan1998}
{Yan}, M., {Sadeghpour}, H.~R., \& {Dalgarno}, A. 1998, \apj, 496, 1044

\bibitem[{{{\.Z}ycki} {et~al.}(1999){{\.Z}ycki}, {Done}, \&
  {Smith}}]{zycki1999}
{{\.Z}ycki}, P.~T., {Done}, C., \& {Smith}, D.~A. 1999, \mnras, 309, 561

\end{thebibliography}

\end{document}